\documentclass{amsart}

\newcommand {\supplus}{\mathop{{\supset}\llap{\raise 
0.5pt\hbox{\normalfont\small+}\hskip 0.5pt}}} 

\newcommand {\subplus}{\mathop{{\subset}\llap{\raise 
0.5pt\hbox{\normalfont\small+}\hskip 0.5pt}}}  

\newcommand {\Cee}    {{\mathbb  C}}

\newcommand {\Kee}    {{\mathbb  K}}

\newcommand {\Zee}    {{\mathbb  Z}}

\newcommand {\fa}     {{\mathfrak{a}}}

\newcommand {\fas}    {{\mathfrak{as}}}
\newcommand {\faut}   {{\mathfrak{aut}}} 
\newcommand {\fb}     {{\mathfrak{b}}}

\newcommand {\fder}   {{\mathfrak{der}}}   %

\newcommand {\ff}     {{\mathfrak{f}}}

\newcommand {\fg}     {{\mathfrak{g}}}    %
\newcommand {\fgl}    {{\mathfrak{gl}}}  %
\newcommand {\fh}     {{\mathfrak{h}}}
\newcommand {\fhei}   {{\mathfrak{hei}}}
\newcommand {\fk}     {{\mathfrak{k}}}

\newcommand {\fL}     {{\mathfrak{L}}}

\newcommand {\fm}     {{\mathfrak{m}}}
\newcommand {\fn}     {{\mathfrak{n}}}

\newcommand {\fo}     {{\mathfrak{o}}}
\newcommand {\fosp}   {{\mathfrak{osp}}}
\newcommand {\fpe}    {{\mathfrak{pe}}}   %
\newcommand {\fpg}    {{\mathfrak{pg}}}
\newcommand {\fpm}    {{\mathfrak{pm}}}
\newcommand {\fpgl}   {{\mathfrak{pgl}}}

\newcommand {\fpsl}   {{\mathfrak{psl}}}
\newcommand {\fpq}    {{\mathfrak{pq}}}

\newcommand {\fpsq}   {{\mathfrak{psq}}}
\newcommand {\fq}     {{\mathfrak{q}}}     
\newcommand {\fr}     {{\mathfrak{r}}}

\newcommand {\fs}     {{\mathfrak{s}}}

\newcommand {\fsg}    {{\mathfrak{sg}}}

\newcommand {\fsi}    {{\mathfrak{si}}}
\newcommand {\fsl}    {{\mathfrak{sl}}}

\newcommand {\fsp}    {{\mathfrak{sp}}}
\newcommand {\fspe}   {{\mathfrak{spe}}}

\newcommand {\fsq}    {{\mathfrak{sq}}}

\newcommand {\fst}    {{\mathfrak{st}}}
\newcommand {\fsvect} {{\mathfrak{svect}}}

\newcommand {\fvect}  {{\mathfrak{vect}}}   %

\newcommand {\cal} {\mathcal}

\newcommand {\cA}     {{\cal A}}
\newcommand {\cB}     {{\cal B}}

\newcommand {\cF}     {{\cal F}}

\def \opname#1#2%
  {\expandafter\newcommand \csname #1\endcsname {{\mathop{#2}\nolimits}}}

\newcommand{\rmname}[1]
  {\expandafter\newcommand \csname #1\endcsname {{\operatorname{#1}}}}

\newcommand{\rmnameii}[2]
  {\expandafter\newcommand \csname #1\endcsname {{\operatorname{#2}}}}

\rmname{act}
\rmname{Ad}
\rmname{Add}
\rmname{ad}
\rmname{Alt}
\rmname{alt}
\rmname{Ann}
\rmname{antidiag}
\rmname{Ber}
\rmname{ber}
\rmname{Br}
\rmname{card}
\rmname{ch}
\rmname{Char}
\rmname{cem}
\rmname{cj}
\rmname{Cliff}
\rmname{cntr}
\rmname{codim}
\rmname{coind}
\rmname{const}
\rmname{col}
\rmname{cork}
\rmname{cpr}
\rmname{diag}
\rmnameii{Div}{div}
\rmname{Def}
\rmname{Der}
\rmname{Dim}
\rmname{End}
\rmname{Even}
\rmname{Ext}
\rmname{gr}
\rmname{Hom}
\rmname{HT}
\rmnameii{Ht}{ht}
\rmname{hwt}
\rmname{Id}
\rmname{id}
\rmname{ind}
\rmname{Ind}
\rmname{Inf}
\rmname{irr}
\rmname{Le}
\rmname{Lie}
\rmname{lwt}
\rmname{mult}
\rmname{Mor}
\rmname{nm}
\rmname{Ob}
\rmname{Odd}
\rmname{Osc}
\rmname{per}
\rmname{Pic}
\rmname{pr}
\rmname{pro}
\rmname{Prime}
\rmname{Proj}
\rmname{prt}
\rmname{pt}
\rmname{Q}
\rmname{qet}
\rmname{qtr}
\rmname{rd}
\rmname{rk}
\rmname{row}
\rmname{Res}
\rmname{salt}
\rmname{Sch}
\rmname{SBr}
\rmname{scalar}
\rmname{Ser}
\rmname{sign}
\rmname{Smbl}
\rmname{spin}
\rmname{ssym}
\rmname{str}
\rmname{st}
\rmname{sgn}
\rmname{sq}
\rmname{symm}
\rmname{supp}
\rmname{Supp}
\rmname{St}
\rmname{Spec}
\rmname{Spm}
\rmname{tr}
\rmname{vpt}
\rmname{weyl}
\rmname{Weyl}
\rmname{Witt}

\opname{vvol}  {{v\hspace{-0.1ex}o\hspace{-0.02ex}l\/}}
\opname{pnt}  {\text{\normalfont pt}}
\opname{Span} {{Span}}
\opname{slim} {\overline{\lim}}
\opname{Vol}  {{V\hspace{-0.55ex}o\hspace{-0.02ex}l\/}}
\opname{QVol} {{Q\hspace{-0.3ex}V\hspace{-0.55ex}o\hspace{-0.02ex}l\/}}
\opname{PoVol}{{P\hspace{-0.35ex}o\hspace{-0.25ex}V\hspace{-0.55ex}o\hspace{-0.02ex}l\/}}
\opname{BVol} {{B\hspace{-0.2ex}V\hspace{-0.55ex}o\hspace{-0.02ex}l\/}}
\opname{Par}  {{P\hspace{-0.3ex}a\hspace{-0.05ex}r\/}}

\rmname{Mat}
\rmname{Bil}
\rmname{Diff}
\rmname{Ker}
\rmname{Herm}
\rmname{Coker}
\rmname{Conn}
\rmname{Covect}
\rmname{Vect}
\rmname{Int}

\rmnameii {IM} {Im}
\rmnameii {RE} {Re}

\opname{Aut} {{A\hspace{-0.2ex}u\hspace{-0.1ex}t\/}}
\opname{GL} {{G\hspace{-0.3ex}L}}
\opname{SL} {{S\hspace{-0.3ex}L}}
\opname{Exp} {{E\hspace{-0.2ex}x\hspace{-0.1ex}p\/}}
\opname{GQ} {{G\hspace{-0.2ex}Q}}
\opname{OSp} {{O\hspace{-0.25ex}S\hspace{-0.15ex}p\/}}
\opname{Out} {{O\hspace{-0.25ex}u\hspace{-0.15ex}t\/}}
\opname{Spp} {{S\hspace{-0.2ex}p\/}}
\opname{SpO} {{S\hspace{-0.2ex}p\hspace{-0.02ex}O\/}}
\opname{Pe} {{P\hspace{-0.25ex}e\/}}
\opname{SPe} {{S\hspace{-0.25ex}P\hspace{-0.25ex}e\/}}
\opname{Spin} {{S\hspace{-0.25ex}p\hspace{-0.05ex}i\hspace{-0.1ex}n\/}}
\opname{Iso} {{I\hspace{-0.25ex}s\hspace{-0.1ex}o\/}}
\opname{SSPe} {{S\hspace{-0.25ex}S\hspace{-0.15ex}P\hspace{-0.25ex}e\/}}
\opname{PeU} {{P\hspace{-0.25ex}e\hspace{-0.1ex}U\/}}
\opname{QU} {{Q\hspace{-0.15ex}U\/}}
\opname{U} {{U\/}}

\opname{cGQ} {{\cal G \hspace{-0.2em} Q \/}}
\opname{cSL} {{\cal S \hspace{-0.2em} L \/}}
\opname{cGL} {{\cal G \hspace{-0.2em} L \/}}
\opname{cOSp} {{\cal O \hspace{-0.2em} S \hspace{-0.3em} \it p\/}}
\opname{cPe} {{\cal P \hspace{-1.5pt} \it e\/}}
\opname{cVect} {{\cal V \hspace{-1.5pt} \it e\hspace{-0.1ex}c\hspace{-0.1ex}t\/}}
\opname{cVol} {{\cal V \hspace{-1.5pt} \it o\hspace{-0.1ex}l\/}}
\opname{cAut} {{\cal A \hspace{-0.2em} \it u\hspace{-0.1em}t\/}}
\opname{cCovect} {{\cal C \hspace{-1.5pt}
     \it o\hspace{-0.1ex}v\hspace{-0.1ex}e\hspace{-0.1ex}c\hspace{-0.1ex}t\/}}
\opname{CW} {{C\hspace{-0.15ex}W}}

\opname {Ab}   {{\sf Ab}}
\opname {Alg}   {{\sf Alg}}
\opname {ASch}  {{\sf Aff\;Sch}}
\opname {Funct}   {{\sf Funct}}
\opname {Gr}   {{\sf Gr}}
\opname {Grf}  {{{\sf Gr}_f}}
\opname {Mods}   {{\sf mods}}
\opname {Rings}   {{\sf Rings}}
\opname {Salg}   {{\sf Salg}}
\opname {Sets} {{\sf Sets}}
\opname {SSMan} {{\sf SMan}}
\opname {Top}  {{\sf Top}}
\opname {Vebun}   {{\sf Vebun}}

\newcommand {\ev} {{\bar0}}
\newcommand {\od} {{\bar1}}

\newcommand {\degree}  {{}^\circ}
\newcommand {\tto} {\longrightarrow}
\newcommand {\pder}[1] {{\frac{\partial}{\partial {#1}}}}
\newcommand {\pderf}[2] {{\frac{\partial {#1}}{\partial {#2}}}}

\newcommand {\bcdot}   {\mathbin{\hbox{\raise.4ex\hbox{\bf.}}}} 

\newcommand {\secno} {}
\newcommand {\ssecfont} {\normalfont\bf}

\newtheorem{Theorem}{\secno Theorem}

\newtheorem{Lemma}[Theorem]{\secno Lemma}

\newenvironment {th*}[1]
    {\gdef\thname{#1} \begin{thn}}%
    {\end{thn}}
\newtheorem{thn}[Theorem] {\thname}

\theoremstyle{definition}

\newenvironment {ex*}[1]
    {\gdef\thname{#1} \begin{exn}}%
    {\end{exn}}
\newtheorem{exn}[Theorem]{\thname}

\theoremstyle{remark}

\newtheorem{Remark}[Theorem]{\secno Remark}

\newenvironment {rem*}[1]
    {\gdef\thname{#1} \begin{remn}}%
    {\end{remn}}
\newtheorem{remn}[Theorem]{\thname}

\newcommand {\ssec}{\subsection*}

\newcommand {\ssbegin}[2]
  {\def \secno {\gdef \secno {}{\ssecfont #1. }}%
   \begin{#2}}
\begin{document}

\title{Maximal subalgebras of matrix Lie superalgebras}

\author{Irina Shchepochkina} 

\address{On leave of absence from the Independent Univ. of Moscow. Correspondence:
c/o D. Leites, Dept. of Math., Univ. of Stockholm  Roslagsv. 101,  Kr\"aftriket hus
6,  S-106 91, Stockholm, Sweden}



\thanks{I am thankful to D. Leites for help; INTAS grant 94-4720 and NFR for
partial financial support and Stockholm University for hospitality.}

\begin{abstract} Dynkin's classification of maximal subalgebras of simple
finite dimensional complex Lie algebras is generalized to matrix Lie
superalgebras, i.e., the Lie subsuperalgebras of $\fgl(p|q)$. 
\end{abstract} 

\maketitle

\section*{Introduction}

\ssec{0.1. Dynkin's result} In 1951 Dynkin published two remarkable papers
[D1],  [D2], somewhat interlaced in their theme: classification
of semisimple subalgebras of simple (finite dimensional)
Lie algebras and classification of maximal subalgebras of
simple (finite dimensional) Lie algebras. These
classifications are of interest {\it per se}; they also proved
to be useful in the studies of integrable systems and in
representation theory.

According to the general ideology of Leites' {\it Seminar on supermanifolds} we
have to generalize Dynkin's results to \lq\lq classical" Lie superalgebras,
i.e., to Lie algebras of polynomial growth \lq\lq close" in a sence to simple ones.
This problem was raised by A.~L.~Onishchik and D.~Leites. In this paper (partly
preprinted long ago [Sh1]) I try to give the answer in a form similar to that of
Dynkin's result.

Let me first remind Dynkin's result. Let $\fg$ be a simple matrix Lie algebra
over $\Cee$, i.e., $\fsl(n)$, $\fo(n)$ or $\fsp(2n)$. In what follows a \lq\lq
matrix" Lie (super)algebra means that the elements of the Lie (super)algebra in
question is realized by matrices. Let
$V$ be the space of the standard (identity) representation
$\id$ of $\fg$ in the space of column-vectors. Let
$\fh\subset\fg$ be a maximal subalgebra. Then only the following 3 cases
might occure:

1) the representation of $\fh$ in $V$ is reducible (we say that $\fh$ is
reducible);

2) $\fh$ is simple and irreducible (i.e., $\fh$ irreducibly acts on $V$);

3) $\fh$ is irreducible but not simple.

The answers in these cases are:

1) $\fh$ can be described as the collection of all operators from $\fg$ that
preserve a subspace $W\subset V$. Here $W$ can be arbitrary for $\fg=\fsl$
case while for $\fg=\fo(n)$ and $\fsp(2n)$ the biliniar form $B$ on $V$
preserved by $\fg$ must be either nondegenerate or identically zero on $W$. 

2) For practically every irreducible representation of a simple $\fh$ the image
is a maximal subalgebra in a simple matrix algebra and Dynkin listed the
exceptions.

3) Here is the complete list of nonsimple {\it maximal irreducible subalgebras}
(abbreviation for {\it maximal subalgebras which are irreducible}):
$$
\begin{matrix} 
&\fsl (V_1)\oplus \fsl (V_2) &\text{ in}&\; \fsl (V_1\otimes V_2) &
\text{ if}& \dim V_2 \ge \dim V_1 \ge 2 \\
&\fsp (V_1)\oplus \fo (V_2) &\text{ in}& 
\fsp (V_1\otimes V_2)\; &\text{ if}& 
\dim V_1\ge 2,\;  \dim V_2\geq 3 \; \\
&&&&&\text{ and}\; \dim V_2\neq 4 \; \text{ or}\\
&&&&\text{ if}& \dim V_1=2 \; \text{ and}\; \dim V_2=4; \\
&\fsp (V_1)\oplus \fsp (V_2)&&&\text{ if}& \dim V_2\ge \dim V_1>2 \\
\text{ or}&&\text{ in}& \fo (V_1\otimes V_2) && \\
&\fo (V_1)\oplus \fo (V_2)&&&\text{ if}& \dim V_2\ge \dim V_1\ge 3\\
&&&&&\text{ and} \; \dim V_1,\, \dim V_2\neq 4. \end{matrix} \eqno{(*)}
$$

\vskip 0.2 cm

Difficulties we encounter in superization of Dynkin's results 1)--3)
range widely:

Superization of 1) to Lie superalgebras with Cartan matrix seem to be
straightforward. Observe, however, that among classical Lie superalgebras, as well
as among infinite dimensional Lie algebras, we naturally encounter algebras which
can not be described with the help of Cartan matrix. Such algebras possess,
nevertheless, a system of roots that can be split into the disjoint union of
positive and negative ones and the notion of a simple root is well-defined.
For a description of systems of simple roots see [PS], [DP]. The
corresponding generators are referred to as {\it Chevalley generators}. In
the absence of Cartan matrix  generators do not quite correspond to the
simple-root vectors, see [LSh1]; they are called {\it generalized Chevalley
generators}. Clearly, {\it parabolic} subalgebras --- distinguished by just
one missing generalized Chevalley generator --- are maximal. Contrary to
the naive expectations they do not exhaust all maximal reducible
subalgebras even in matrix Lie superalgebras considered in [ZZO], where a
minor mistake of [D2] is corrected in passing. We will consider this case
elsewhere [LSh2].

Superization of 2) was impossible untill recently. For a partail result see
[J1]. Now that Penkov and Serganova [PS] derived the character formula for
any irreducible finite dimensional representation of a finite dimensional
Lie superalgebra we can consider this problem and intend to do it
elsewhere. 

Superization of 3) is what is done in this paper: the description
the irreducible nonsimple maximal subsuperalgebras of matrix Lie superalgebras
either simple or \lq\lq classical" --- certain algebras closely related to
simple ones.

Regrettably, a precise description of the maximal Lie superalgebras of type 3)
is more involved than (already not very concise) Dynkin's description $(*)$
above. Therefore, in order to grasp the formulation, we split it into several
statements (Theorems 0.2.1--0.2.6). Still, a sufficiently transparent and
comparable with $(*)$ Main Theorem (below) singles out a small amount of
contenders for the role of maximal subalgebras; these contenders almost always
are, indeed, maximal.

To make the paper readable we recall the list
of classical matrix Lie superalgebras and the general background on superspaces
(see Appendix 2). Assuming the reader is familiar with these notions, let us
formulate the result explicitely. 

{\bf 0.2. Our result}. Throughout the paper the ground field is $\Cee$. In
order to formulate our result, we need several notations listed under bullets. 

$\bullet$ Given two representations of Lie superalgebras $\fg_i\tto\fgl(V_i)$ we
get a representation of the Lie superalgebra $\fg_1\oplus\fg_2\tto\fgl(V_1\otimes
V_2)$ in the superspace $V=V_1\otimes V_2$ , called the {\it tensor product} of
the given representations, setting 
$$
g_1+g_2\mapsto g_1\otimes 1+1\otimes g_2\quad \text{for }g_1\in\fg_1, \;
g_2\in\fg_2. \eqno{(0.2)}
$$
If both $\fg_1$ and $\fg_2$ contain the identity operators, the representation
$(0.2)$ has a 1-dimensional kernel. By $\fg_1\bigodot \fg_2$ we will mean the
image of the direct sum under the representation $(0.2)$ if not both the $\fg_i$
are of type $\fq$. It is convenient to retain the notation $\fg_1\bigodot
\fg_2$ even in the absence of the kernel, i.e., when $\fg_1\bigodot
\fg_2=\fg_1\oplus\fg_2$.

$\bullet$ If both the $\fg_i$ are of type $\fq$ the construction of 
$\fg_1\bigodot
\fg_2$ is different. Namely, let $V_0$ be a $(1, 1)$-dimensional superspace and
$\cA \subset \Mat (1|1)=\Mat (V_0)$ the associative superalgebra of matrices
of the form $\begin{pmatrix}a & b \\ b & a\end{pmatrix}$, where $a, b\in \Cee $.
Then the {\it associative} superalgebra $C(\cA )$ --- the centralizer of
$\cA $ --- consists of matrices of the form
$\begin{pmatrix} c & d \\ -d & c\end{pmatrix}$, where
$c, d\in \Cee $. Clearly, $\cA \cong C(\cA )$.

Observe that the Lie superalgebra $\fq (n)$ can be
represented as the tensor product $\fgl (n|0)\otimes \cA $ of the 
Lie algebra $\fgl (n|0)$ by the associative superalgebra $\cA $. Then
the space $W$ of the standard representation of $\fq (n)$ can be
renresented as $W=W_{\bar 0}\otimes \cA $ in a natural way:
$$
(g\otimes a_{1})(w\otimes a_{2})=g\cdot w\otimes a_{1}a_{2}\; \text{ for any}\;
g\in \fgl (n|0), w\in W_{\bar 0}, a_{1}, a_{2}\in \cA .
$$

Let now a superspace $V$ be represented as $V=V_{1}\otimes \cA \otimes 
V_{2}$,  where $\dim V_{1}=(n_{1}, 0), \dim V_{2}=(n_{2}, 0)$. Then a natural 
action $\rho$ of $\fq(n_{1})\oplus \fq (n_{2})$ is determined in 
$V$: indeed,  $\fq (n_{1})$ acts in $V_{1}\otimes \cA $, and 
$\fq (n_{2})$ acts in $\cA \otimes V_{2}\cong C(\cA )\otimes V_{2}$.

In terms of matrices this action takes the form:
$$
\begin{matrix} 
\begin{pmatrix} A & B \\ B & A \end{pmatrix} \in \fq (n_{1}) & 
\mapsto \begin{pmatrix} A\otimes 1 & B\otimes 1 \\ B\otimes 1 & A\otimes
1\end{pmatrix}\subset\fgl(V_{1}\otimes \cA ) ; \\
\begin{pmatrix} C & D \\ D & C\end{pmatrix} \in \fq (n_{2}) & 
\mapsto \begin{pmatrix} 1\otimes C & i\otimes D \\ -i\otimes D & 1\otimes
C\end{pmatrix} \subset\fgl(\cA \otimes V_{2}) .
\end{matrix} 
$$
The representation $\rho$ has a one-dimensional kernel. Denote the quotient of 
$\fq(n_{1})\oplus \fq (n_{2})$ modulo this kernel by $\fq(n_{1})\bigodot\fq
(n_{2})$.

$\bullet$ We often use a general notation $\faut (B)$ for the Lie superalgebra
that preserves the bilinear form $B$.  The Lie superalgebra $\faut (B)$ turns into
either $\fosp$ or $\fpe$ depending on the parity of $B$.

$\bullet$ Denote by $\fhei(0|2n)$ the Heisenberg Lie superalgebra with $2n$ odd
generators of creation and annihilation, i.e., the Lie superalgebra with 
odd generators $\xi_1$, \dots , $\xi_n$;  $\eta_1$, \dots , $\eta_n$ and an
even generator $z$ satisfying the relations
$$
[\xi_i, \eta_j]=\delta_{i, j}\cdot z, \; [\xi_i, \xi_j]=[\eta_i, \eta_j]= [z,
\fhei(0|2n)]=0.
$$
The only irreducible finite dimensional representation of $\fhei(0|2n)$ that
sends $z$ into the identity operator is realized on the superspace
$\Lambda(n)=\Lambda (\xi)$ by the formulas $\xi_i \mapsto \xi_i$, $\eta_i\mapsto
\partial_i=\pder{\xi_{i}}$. 

The normalizer of $\fhei(0|2n)$ in $\fgl(\Lambda (n))$ is $\fg=\fhei(0|2n)
\supplus\fo(2n)$; we can realize $\fg$ in the space of the spinor representation
of $\fo(2n)$ by the following differential operators: $\fg=\Span(1; \xi,
\partial; 
\xi\xi,  \xi\partial, \partial\partial)$.

$\bullet$ Define the representation $T^{1/2}$
of
$\fvect(0|n)$ in the superspace $\Lambda (\xi)\sqrt{\vvol(\xi)}$ of halfdensities
by the formula
$T^{1/2}(D)=D+\frac 12\Div D$ and define the form $\omega$ on $\Lambda
(\xi)\sqrt{\vvol}$ by the formula
$\omega(f\sqrt{\vvol}, g\sqrt{\vvol})=\int fg\cdot \vvol(\xi)$.

\begin{th*}{Main Theorem} $1\degree$ Let $\fg$ be an irreducible matrix
Lie superalgebra which is neither almost simple nor a central
extension of an almost simple Lie superalgebra.

Then $\fg$ is contained in one of the following four major types of Lie
superalgebras and only in them:

{\em 1)} $\fgl(V_1)\bigodot \fgl(V_2)$;

{\em 2)} $\fq(V_1)\bigodot \fq(V_2)$;

{\em 3)} $\fgl(V)\otimes \Lambda(n)\supplus \fvect(0|n)$;

{\em 4)} $\fhei(0|2n)\supplus \fo(2n)$.

\noindent $2\degree$ Let in addition to conditions of $1\degree$, $\fg$ be a
subalgebra of
$\fq(V)=C(J)$. Then $\fg$ is contained in one of the following Lie superalgebras
(numbered as in $1\degree$) and only in them:

$1\fq)$ $\fq(V_1)\bigodot \fgl(V_2)$ and $J=J_1\otimes 1$, where
$\fq(V_1)=C(J_1)$;

$3\fq)$ $\fq(V_0)\otimes \Lambda(n)\supplus \fvect(0|n)$ and $J=J_0\otimes 1$, where
$\fq(V_0)=C(J_0)$;

\noindent $3\degree$ Let, in addition to conditions of $1\degree$, $\fg$ preserve
a nondegenerate homogeneous form $\omega$, either symmetric or skew-symmetric.
Then $\fg$ is contained in one of the following Lie superalgebras
(numbered as in $1\degree$) and only in them:

$1\omega)$ $\faut(\omega_1)\bigodot \faut(\omega_2)$ and $\omega=\omega_1\otimes
\omega_2$;

$3\omega)$ $T^{1/2}(\fvect(0|n))$, where $\omega$ is the form on
half-densities. 
\end{th*}

The union of the following statements 0.2.1--0.2.6 constitute, more or less, a
statement converse to Main Theorem. The basic types of examples considered in
these theorems appear in \S 1 labelled 1.1 -- 1.4. 

\ssbegin{0.2.1}{Theorem} Let $\dim V_i=(n_i, m_i)$ and $(n_i, m_i)\neq (1,
1)$,
$(1, 0)$, $(0, 1)$. Then

$1)$ $\fgl(V_1)\bigodot \fgl(V_2)$ is maximal irreducible in $\fgl(V)$
if either $n_1\neq m_1$ or $n_2\neq m_2$;

$2)$ the superalgebras $\fsl(V_1)\oplus \fsl(V_2)$ if either $n_1\neq m_1$ or
$n_2\neq m_2$ and $\fgl(V_1)\bigodot \fgl(V_2)$ if $n_1=m_1$ and $n_2=m_2$
are maximal irreducible in $\fsl(V)$.
\end{Theorem}

\ssec{0.2.2} Let $\dim V_1=(m, m)$; let $V_2$ be of any (finite)
superdimension. For odd operator $J_1\in\End (V_1)$ such that $J_1^2=-1_{V_{1}}$,
we clearly see that $J=J_1\otimes 1$ satisfies $J^2=-1_{V}$. Set
$\fq(V_1)=C(J_1)$ and 
 $\fq(V)=C(J)$.

\begin{Theorem} The Lie superalgebra $\fq(V_1)\bigodot
\fgl(V_2)$ is maximal irreducible in $\fq(V)$.
\end{Theorem}

\ssec{0.2.3} Let $\fg_1=\fq(V_1)=\fq(n_1)=C(J_1)$ and
$\fg_2=\fq(V_2)=\fq(n_2)=C(J_2)$. The representation $\rho$ of $\fg_1\oplus
\fg_2$ in $V=V_1\otimes V_2$ has, clealry, the $2|2$-dimensional Lie superalgebra
of intertwining operators spanned by
$$
\id=1_V, \; \; J_1\otimes 1, \; \; 1\otimes J_2, \; \; J_1\otimes J_2.
$$
Set $J=J_1\otimes J_2$. Clearly, $J^2=-1_V$; the restrictions $\rho_{\pm i}$
of $\rho$ to the eigenspaces $V_{\pm i}$ of $J$ are irreducible and the change
of parity functor $\Pi$ permutes $V_{i}$ with $V_{-i}$.

\begin{th*}{Theorem} The images $\rho_{\pm i}(\fq(V_1)\oplus \fq(V_2))$ are
isomorphic to $\fq(V_1)\bigodot \fq(V_2)$. They are maximal irreducible in
$\fsl(V_{\pm i})$. 
\end{th*}

\ssec{0.2.4} Let us consider Lie superalgebras $\fg_i=\faut (\omega_i)$ each
preserving a bilinear form
$\omega_i$ in the superspace $V_i$; let $B_i$ be the matrices of
these forms, $i=1$, 2. Clearly, in $V=V_1\otimes V_2$ the form
$\omega=\omega_1\otimes \omega_2$ of parity $p(\omega_1)+p(\omega_2)$ arises; 
if the $\omega_i$ are nondegenerate, then so is $\omega$. Let
$\fg=\faut(\omega)$. Formula $(0.2)$ defines a representation of $\fg_1\oplus
\fg_2$ in $\fg$.

\begin{th*}{Theorem} The Lie superalgebra $\fg_1\oplus \fg_2$ is maximal
irreducible in $\fg$ if $p(\omega)=\ev$ or if $p(\omega_1)=\ev$,
$p(\omega_2)=\od$ and $\dim V_1\neq (n, n)$. 

If $p(\omega_1)=\ev$, $p(\omega_2)=\od$ and $\dim V_1=(n, n)$, then $\fg_1\oplus \fg_2$ is maximal
irreducible in $\fg\cap \fsl(V)$.
\end{th*}

More exactly, the following subalgebras $\fg _{1}(V_{1})\oplus \fg 
_{2}(V_{2})$ are maximal in $\fg (V_{1}\otimes V_{2})$:
$$
\begin{array}{|c|c|c|}
\hline
\fg _{1} & \fg _{2} & \fg \\
\hline
\fosp (n_{1}|2m_{1}) & \fosp (n_{2}|2m_{2}) & 
\fosp (n_{1}n_{2}+4m_{1}m_{2}|2n_{1}m_{2}+2n_{2}m_{1}) \\
\fo (n) & \fosp (n_{2}|2m_{2}) & \fosp (nn_{2}
|2nm_{2}), n\neq 2, 4 \\
\fsp (2n) & \fosp (n_{2}|2m_{2}) & 
\fosp (2mn_{2}|4nm_{2}) \\
\fpe (n_{1}) & \fpe (n_{2}) & 
\fosp (2n_{1}n_{2}|2n_{1}n_{2}), n_{1}, n_{2}>2 \\
\hline
\fosp (n_{1}|2m_{1}) & \fpe (n_{2}) & 
\fpe (n_{1}n_{2}+2m_{1}n_{2})\; \text{ if}\; n_{1}\neq 2m_{1} \\
& & \fspe (n_{1}n_{2}+2m_{1}n_{2})\; \text{ if}\; n_{1}=2m_{1} \\
\fo (n) & \fpe (m) & \fpe (nm) \\
\fsp (2n) & \fpe (m) & \fpe (2nm) \\
\hline
\end{array}
$$

\ssec{0.2.5} The maximal subalgebras considered in sec. 0.2.1--0.2.4 are similar
to those considered by Dynkin. There are, however, maximal subalgebras of matrix
superalgebras of totally different nature.

Let $V_1$ be a linear superspace of dimension $(r, s)$; let $\Lambda (n)$ the
Grassmann superalgebra with $n$ odd generators $\xi_1$, \dots $\xi_n$ and
$\fvect(0|n)=\fder \Lambda (n)$ the Lie superalgebra of vector fields on the
$(0, n)$-dimensional supermanifold.

Let $\fg=\fgl(V_1)\otimes \Lambda (n)\supplus \fvect(0|n)$ be the semidirect
product with the natural action of $\fvect(0|n)$ on the ideal $\fgl(V_1)\otimes \Lambda
(n)$. The Lie superalgebra $\fg$ has a natural faithful representation $\rho$
in the space $V=V_1\otimes \Lambda (n)$ defined by the formulas
$$
\begin{matrix} 
\rho(X\otimes \varphi)(v\otimes\psi)=(-1)^{p(\varphi)p(\psi)}Xv\otimes
\varphi\psi;\\
\rho(D)(v\otimes\psi)=-(-1)^{p(D)p(v)}v\otimes
D\psi;\end{matrix}
$$
for any $X\in\fgl(V_1)$, $\varphi, \psi\in  \Lambda (n)$, $v\in V_1$,
$D\in\fvect(0|n)$. 

Let us identify the elements from $\fg$ with their images under $\rho$, so we
consider $\fg$ embedded into $\fgl(V)$.

\begin{th*}{Theorem} $1)$ The Lie superalgebra $\fgl(V_1)\otimes \Lambda
(n)\supplus
\fvect(0|n)$ is maximal irreducible in $\fsl(V_1\otimes \Lambda (n))$ unless
a) $\dim V_1=(1, 1)$ or b) $n=1$ and $\dim V_1=(1, 0)$ or $(0, 1)$ or $(r,
s)$ for $r\neq s$. 

$2)$ If $\dim V_1=(1, 1)$, then
$\fgl(1|1)\cong \Lambda (1)\supplus
\fvect(0|1)$, so  $\fgl(V_1)\otimes \Lambda (n)\supplus
\fvect(0|n)\subset \Lambda (n+1)\supplus
\fvect(0|n+1)$  and it is the latter superalgebra which is maximal irreducible
in $\fsl(V)$. 

$3)$ If $n=1$ and $\dim V_1= (r,s)$ for $r> s>0$, then
$\fg$ is maximal irreducible in $\fgl(V)$. 
\end{th*}

\ssec{0.2.5$'$} Let $\fg_1=\fq(n_1)=\fq(V_1)=C(J_1)$; let $V\cong
V_1\otimes\Lambda (n)$ and $J=J_1\otimes \id_{\Lambda(n)}$. Clearly,
$J^2=-\id_V$. It is also clear that 
$$
\fg_1\otimes \Lambda (n)\supplus\fvect(0|n)\subset \fq(V)=C(J).
$$

\begin{Theorem} The Lie superalgebra $\fq(n_1)\otimes
\Lambda(n)\supplus\fvect(0|n)$ is maximal irreducible in $\fq(V_1\otimes \Lambda
(n))$.
\end{Theorem}

\ssbegin{0.2.6}{Theorem} $1)$ For $n\geq 2$ and $n\neq 3$ the algebra
$\fg=\fhei(0|2n)\supplus\fo(2n)$ is
maximal irreducible in $\fsl(\Lambda (n))$.

$2)$ For $n=3$ the algebra $\fg$ is contained in the nontrivial central
extension $\fas$ of $\fspe(4)$; the Lie superalgebra $\fas$ is maximal irreducible
in $\fsl(\Lambda (3))$.  
\end{Theorem}

\ssec{0.3. Realted results}

\ssec{0.3.1. Comparison with the case of prime characteristic} 
Our Theorems 0.2.5 and 0.2.6 are very similar to the result  in prime
characteristic announced by A.~I.~Kostrikin's student O.~K.~Ten. 

Recall that a subalgebra of a (finite dimensional) Lie algebra
is called {\it regular} if it is invariant with respect to
a maximal torus.

\begin{Theorem}{\em ([T])} Let $\Kee$ be an algebraically closed
field of characteristic $p>3$.

$1)$ Any non-semisimple maximal subalgebra
of $\fsl (n)$ for $n\not\equiv 0 \mod p$, $\fo (n)$ or 
$\fsp (2n)$ is regular.

{\em 2a)} Let $\dim V=np^{m}$, $(n, p)=1$, $n>1$. If $\fm$ is an 
irreducible maximal subalgebra in $\fsl (V)$ such that $\fpm =\fm /<1>$ is not 
semisimple, then $V=U\otimes O_{m}$, where $O_{m}=\Kee [x_{1}, \dots ,
x_{m}]/(x^{p}_{1},  \dots , x^{p}_{m})$ and $\fm =\fgl (U)\otimes
O_{m}\supplus \fvect _{m}$. 

{\em 2b)} If $n=1$, then in addition to the above examples $2a)$ the algebra 
$\fm =\fsp _{2m}\supplus \fhei (2m|0)$ is also maximal in 
$\fsl (V)$. 

$3)$ Any maximal subalgebra in $\fg _{2}$ is regular except $\fvect _{1}$ for
$p=7$  and $\fsl (2)$ for $p>7$. 
\end{Theorem}

\ssec{0.3.2. The maximal (almost) simple subalgebras of simple finite
dimensional Lie superalgebras} It seems that it is only possible to describe
such subalgebras when they are simple and, moreover, are isomorphic to $\fosp
(1|2n)$, whose finite dimensional representations are completely reducible.
Perhaps, it is possible to extend the result to
$\fosp (2|2n)$ and $\fsl (1|n)$, whose finite dimensional representations are \lq\lq
tame" in a sence (see [L4]) and, therefore, describable.

$\bullet$ Embeddings of $\fosp (1|2)$ are described by Serganova
in [LSS]. (Partly, her results were independently rediscovered by
J.~Van~der~Jeugt [J2], where {\it regular} Lie subsuperalgebras of finite
dimensional simple Lie superalgebras with Cartan matrix are discribed, cf.
[LSh2].)

$\bullet$ A subalgebra $\fh$ of the Lie superalgebra $\fg$ is called {\it
Volichenko algebra} if it is not a Lie subsuperalgebra. A list of simple
finite dimensional Volichenko subalgebras in simple Lie superalgebras is
obtained under a technical condition by Serganova [S]. (For motivations and
infinite dimensional case see [LS] and [KL].)

\ssec{0.3.3. Nonsimple maximal subalgebras} 
Maximal {\it solvable} Lie superalgebras of $\fgl(m|n)$ and $\fsl(m|n)$ are
classified in [Sh2]. A strange series of subalgebras was discovered.

\ssec{0.3.4. Maximal subalgebras of nonmatrix superalgebras} In the case when
the ambient possesses a Cartan matrix this is done in [J2] with a gap filled in
by Serganova (published in [Sh3]). For the classification of graded maximal
subalgebras of vectoral Lie superalgebras see [Sh1] and [Sh3].

\section*[\S 1. Irreducible non-simple matrix maximal subalgebras]
{\protect\S 1. Irreducible non-simple matrix maximal subalgebras\\ of
$\fgl(m|n)$ and $\fsl(m|n)$}

The 4 basic types of examples of maximal subalgebras from Main
Theorem are considered in 1.1 -- 1.4.

\ssbegin{1.1}{Theorem} Let $\dim V_{i}=(n_{i}, m_{i})$ and $(n_{i},
m_{i})\neq (1, 1),  (1, 0), (0, 1)$. Let $V=V_{1}\otimes V_{2}$. Then 

$1)$ $\fgl (V_{1})\bigodot \fgl (V_{2})$ is a maximal subalgebra in $\fgl (V)$ if
$n_{1}\neq m_{1}$ or $n_{2}\neq m_{2}$; it is maximal in $\fsl(V)$ otherwise. 

$2)$ The following subalgebras are maximal in $\fsl (V)$: {\em a)} $\fsl
(V_{1})\oplus \fsl (V_{2})$ if $n_{1}\neq m_{1}$ or $n_{2}\neq m_{2}$;
{\em b)} $\fgl (V_{1})\bigodot \fgl(V_{2})$ if $n_{1}=m_{1}$ and 
$n_{2}=m_{2}$.
\end{Theorem}

\begin{proof} 1) The space of all operators in an arbitrary
superspace $V$ may be endowed with two structures: that of
the Lie superalgebra $\fgl (V)$ and of the associative
superalgebra $\Mat (V)$. Observe that 
$\Mat (V_{1})\otimes \Mat (V_{2})\cong\Mat(V_{1}\otimes V_{2})$.
Therefore, any element $g\in \fgl (V_{1}\otimes V_{2})$ can be represented 
in the form $g=\sum A_{i}\otimes B_{i}$, where $A_{i}\in 
\fgl (V_{1})$, $B_{i}\in \fgl (V_{2})$. The bracket in $\fgl(V_{1}\otimes
V_{2})$ is defined via 
$$
[A\otimes B, C\otimes D]=(-1)^{p(B)p(C)}[A, C]\otimes BD+(-1)^{p(A\otimes 
B)p(C)}CA \otimes [B, D]. \eqno{(1.1)}
$$
Now let
$$
\fg =\fgl (V_{1})\bigodot\fgl (V_{2})=\{A\otimes 1+1\otimes B\}, \;
\text{ where}\; A\in \fgl (V_{1}), B\in \fgl (V_{2}),
$$
and let $\ff$ be an intermediate subalgebra, i.e., 
$$
\fg \subset \ff\subset \fgl (V)
\eqno{(1.2)}
$$
Let $h\subset \ff $ be an element which is not contained in $\fg$. Then
$h=\sum A_{i}\otimes B_{i}$, where $A_{i}\in \fgl (V_{1}), B_{i}\in 
\fgl (V_{2})$. Thanks to (1.2) $\ff$ contains all
elements of the form $[g, h], g\in \fg$. Therefore, due to (1.1), we
have
$$
[A\otimes 1, h]=\sum [A, A_{i}]\otimes B_{i}, \; \; \; [1\otimes B, h]=\sum 
(-1)^{p(B)p(A_{i})}A_{i}\otimes [B, B_{i}].
$$
Since $A$ and $B$ are arbitrary operators of $\fgl (V_{1})$ and 
$\fgl (V_{2})$ respectively, $\ff$ contains $h$ and all the elements of
the form $h_{C}=\sum C_{i}\otimes B_{i}$ and $h_{D}=\sum A_{i}\otimes D_{i}$, 
where $C_{i}$ (resp. $D_{i})$ though related perhaps, run over an ideal
$I_{i}\subset \fgl (V_{1})$ (resp.  $J_{i}\subset \fgl (V_{2}))$ for
each fixed $i$. 

Now, observe that, 

1) $\fgl (W)$ contains only two ideals, namely,
$I_{0}=\Cee \cdot 1$ and $I_{1}=\fsl (W)$ for any $W$ of dimension not equal
to $(1, 1)$, (1, 0), (0, 1) and

2) $\ff $ contains the subalgebras
$\fgl (V_{1})\otimes 1$ and $1\otimes \fgl (V_{2})$.

This implies that $\ff$ contains the set $\fsl (V_{1})\otimes \fsl (V_{2})$.
Since the set $\fsl(W)$ is not an associative subalgebra of $\Mat (W)$,
then (1.1) implies that 
$$ 
\ff \supset \fsl (V_{1})\otimes \Mat (V_{2})\oplus \Mat
(V_{1}) \otimes \fsl (V_{2})=\fsl (V_{1}\otimes V_{2}).
$$
If, in addition, $n_{1}\neq m_{1}$, then $\str (1\otimes B)=(n_{1}-m_{1})
\str B\neq 0$, if $\str B\neq 0$, i.e., $\ff$ coincides with the whole $\fgl
(V)$. 

If $n_{1}=m_{1}$ and $n_{2}=m_{2}$, then $\fgl (V_{1})\bigodot\fgl (V_{2})$ is
contained in $\fsl (V)$ and by the above arguments is a maximal Lie
subsuperalgebra of $\fsl (V)$. 

To conclude the proof of heading 2a), it only suffices to notice that
$1\otimes B\in \fsl (V)$ if and only if $B\in \fsl (V_{2})$ for $n_{1}\neq
m_{1}$ and $B\in \fgl (V_{2})$ for $n_{1}=m_{1}$. \end{proof}

\ssec{1.2}Let  $W_{1}$ and $W_{2}$ be standard
representations of $\fg _{1}=\fq (n)=C(J_{1})$ and $\fg _{2}= \fq
(m)=C(J_{2})$, respectively. The representation $\rho $ of $\fg =\fg
_{1}\oplus \fg _{2}$ in the tensor product $V=V_{1}\otimes V_{2}$ has the $(2,
2)$-dimensional algebra of intertwining operator $<1\otimes 1, J_{1}\otimes 1,
1\otimes J_{2}, J_{1}\otimes J_{2}>$. The even intertwining operator
$J=J_{1}\otimes J_{2}$ satisfies $J^{2}=-1$. The restrictions $\rho _{\pm i}$
of $\rho $ onto the eigensubspaces $V_{i}$ and $V_{-i}$ corresponding to
eigenvalues $\pm i$ of $J$ are irreducible and are transformad from each other
by the change of parity functor.

It is easy to verify that given $Q$-bases
$$
(v_{1}, \dots , v_{n}, J_{1}v_{1}, \dots , J_{1}v_{n}) \text{ and $(w_{1}, 
\dots , w_{n}, J_{2}w_{2}, \dots , J_{2}w_{n})$}
$$
of the spaces $V_{1}$ and $V_{2}$ respectively, the basis of $V_{\pm i}$ is the
form
$$
iv_{k}\otimes w_{l}\pm J_{1}v_{k}\otimes J_{2}w_{l}, v_{r}\otimes J_{2}w_{s}+
iJ_{1}v_{r}\otimes w_{s}; \quad k, r=1, \dots , n;\quad l, s=1, \dots , m.
$$
Clearly, $\dim V_{\pm i}=(nm, nm)$ and the homogeneous subspaces of
$V_{\pm i}$ may be identified with the tensor products
$$
(V_{\pm i})_{\bar 0}=(V_{1})_{\bar 0}\otimes (V_{2})_{\bar 0}, \quad 
(V_{\pm i})_{\bar 1}=(V_{\bar 1})\otimes (V_{2})_{\bar 0}.
$$
Let $(A, B)\in \fg _{1}$ and $(C, D)\in \fg _{2}$, then
$$
\rho _{\pm i}((A, B))=(A\otimes 1, B\otimes 1),\quad \rho _{\pm i}((C, D))=
(1\otimes C, 1\otimes (\pm iD))
\eqno{(1.3)}
$$

\ssbegin{1.2.1}{Theorem} The image $\rho _{\pm i}(\fg )$ is a maximal 
subalgebra of $\fsl (V_{\pm i})$.
\end{Theorem}

\begin{proof} It is clear from (1.3), that the restriction of $\rho _{i}$
to $\ff _{1}=\fq (n_1)\oplus \fq (n_2)_{\bar 0}\equiv\fq (n_1)\oplus
\fgl(n_2)$ is the tensor product of the standard representations of $\ff_{1}$ in
$V_{1}$ and $\fq (n_2)_{\bar 0}=\fgl (n_2)$ in $(V_{2})_{\bar 0}$. As
follows from the proved above, $\ff _{1}$ is a maximal subalgebra of the Lie
superalgebra $\fa  _{1}=C(J_{1}\otimes 1)=
\left\{ 
\begin{pmatrix}A& B \\ 
B & A\end{pmatrix}\right\}\subset\fgl (nm|nm)$.

Similarly, $\rho _{i}|\ff _{2}$ for $\ff _{2}=
\fq (n)_{\bar 0}\oplus \fg _{2}$ is the tensor product of
the standard representation of $\fq (n)_{\bar 0}=
\fgl (n)$ in $(V_{1})_{\bar 0}$ and $\fq (m)$
in $V_{2}$. Therefore, $\ff _{2}$ is a maximal subalgebra in the Lie
superalgebra
$$
\fa  _{2}=C(1\otimes J_{2})=\left\{ \begin{pmatrix}A & B \\ -B & A\end{pmatrix} 
\right\} 
\subset \fgl (nm|nm).
$$
Since $\fa  _{1}$ and $\fa  _{2}$ genetare the whole $\fsl (V_{i})$, it
follows that $\fg $ is maximal in $\fsl (V_{i})$. 
\end{proof}

Let us formulate this in another way.

\ssbegin{1.2.2}{Theorem} The Lie subsuperalgebra $\fg=\fq (V_{1}\otimes {\cal
A})\bigodot \fq (\cA \otimes V_{2})$ is maximal in $\fsl (V_{1}\otimes \cA \otimes 
V_{2})$. 
\end{Theorem}

\ssec{1.3} Let $\dim V_{1}=(r,s)$, $\fn= \fgl (V_{1})\otimes \Lambda
(n)$ and $\fg$ the semidirect sum of the ideal $\fn$ and the subalgebra
$\fvect (0|n)=\fk$ with the natural action on the ideal. The Lie superalgebra
$\fg$ has a natural faithful representation $\rho$ in $V=V_{1}\otimes \Lambda
(n)$ defined by the formulas 
$$
\begin{matrix} 
\rho (A\otimes \varphi )(v\otimes \psi )&= &(-1)^{p(D)p(v)}Av\otimes \varphi
\psi\\
\rho (D)(v\otimes \psi )&=&-(-1)^{p(D)p(v)}v\otimes D(\psi )\end{matrix}  
\; \text{for}\; A\otimes \varphi \in \fn,\;  D\in \fk,\;  v\otimes \psi\in V. 
 $$
In the sequel, we will always identify elements of $\fg$
with their images under $\rho $. Therefore, we will consider $\fg$
embedded in $\fgl (V)$ which coincides, as a set, with $\Mat (V)\cong 
\Mat (V_{1})\otimes \Mat (\Lambda (n))$. 

Finally, let us realize the associative superalgebra $\Mat (\Lambda (n))$ as
the associative superalgebra $\Diff (n):=\Diff (0|n)$ of
differential operators acting on the superspace of functions $\Lambda (n)$ on
${\cal C}^{0, n} $.
 
Set $\gr \Diff (n)=\oplus _{0\le i\le n}\Diff_{i}(n)$, where
$\Diff_{i}(n) $ is the superspace of homogeneous (w.r.t. order) order $i$
differential operators. Observe that $\Diff_{0}(n)=\Lambda (n) $, and
$(\Diff_{1}(n))_{L}\cong\fvect (0|n)$. In such a realization the bracket in $\fgl
(V)$ is of the form
$$
[A_{1}\otimes D_{1}, A_{2}\otimes D_{2}]=(-1)^{p(D_{1})p(A_{2})}([A_{1}, A_{2}]
\otimes D_{1}D_{2}+(-1)^{p(A_1)p(A_2)}A_{2}A_{1}\otimes [D_{1}, D_{2}]).
\eqno{(1.4)}
$$

Notice that the subset of matrices of the form $\begin{pmatrix} 
\alpha & \beta \\ 0 & \alpha\end{pmatrix}$
in $\Mat (1|1)$ is, on the one hand, an associative subalgebra
of $\Mat(1|1)$ and, on the other hand, is a commutative ideal
in $\fgl (1|1)$. Therefore, thanks to (1.4)
$$
\tilde \fg =\fgl (1|1)\otimes \Lambda (n)+\left\{ \begin{pmatrix}\alpha &
\beta\\ 0 & \alpha\end{pmatrix}\right\}\otimes \Diff_{1}(n) 
$$
is a Lie superalgebra. Clearly, $\fgl (1|1)=\Lambda (1)\supplus  \fvect (0|1)$ and
$$
\tilde \fg =[\Lambda (1)\supplus  \fvect (0|1)]\otimes \Lambda(n)\oplus \Lambda
(1)\otimes \Diff_{1}(n)\cong \Lambda (n+1)\supplus  \fvect (0|n+1).
$$

\ssbegin{1.3.1}{Theorem} 1) In all cases except a) $\dim V_{1}=(1, 1)$
or b) $n=1$ and $\dim V_{1}=(1, 0)$ or $(0, 1)$ or $(r, s)$ with $r\neq s$
the Lie superalgebra $\fg =\fgl (V_{1})\otimes \Lambda (n)\supplus  
\fvect (0|n)$ is a maximal Lie
subsuperalgebra of $\fsl (V)$,  where $V=V_{1}\otimes \Lambda (n)$.

$2)$ If $\dim V_{1}=(1, 1) $, then $\fg \subset \tilde \fg $ and 
$\tilde \fg $ is a maximal Lie subsuperalgebra
in $\fsl (V) $.

$3) $ If $n=1 $ and $\dim V_{1}=(1, 0) $ or $(0, 1) $, then
$\fg =\fgl (V) $.

$4) $ If $n=1 $ and $\dim V_{1}=(r, s) $, where $r\neq s $ and $r+s>1$,
then $\fg $ is maximal Lie subsuperalgebra of $\fgl (V) $. The Lie
superalgebra $\fsg:=\fgl (V_{1})\otimes \Lambda (1)\supplus  <\partial > $ 
is maximal in $\fsl (V) $.
\end{Theorem}

\begin{Remark} If $\dim V_{1}=(1, 0), (0, 1) $ or $(1, 1) $, then 
 $\fg $ (or  $\tilde \fg $) is a Lie superalgebra of a general form (not
semisimple or likewise). In other cases $\fg $ is semisimple.
\end{Remark}
 
Set $\widetilde{\Diff}_{n}(n)=\{\varphi \partial _{1}\dots
\partial _{n}|\deg \varphi <n $ in the standard grading
of $\Lambda (n)\} $.
 
\ssbegin{1.3.2}{Lemma} {\em 1)} $\fsl (\Lambda (n))\cong \oplus _{0\le i\le
n-1}
\Diff_{i}(n)\oplus \widetilde{\Diff}_{n}(n) $ (as
superspaces).

{\em 2)} The representation of $\fvect (0|n) $ in $\Diff_{i}(n) $
is irreducible for $n>2 $ if $i<n $ and contains the
irreducible subsuperspace $\widetilde{\Diff}_{n}(n) $ of codimension 
$(1, 0) $ if $i=n $.
\end{Lemma}
 
\begin{proof} 1) Let $F=\varphi \partial _{i_{1}}\dots \partial _{i_{k}}\in
\Diff (n)$ and $\deg \varphi =l$. Then $\deg F=l-k$. Hence, if $k\neq
l$, then $F$ is a nilpotent operator and $\str F=0$. If $k=l$, then
$F(\Lambda (n))\subset \varphi \Lambda (n)$. If, moreover,  $\partial
_{i_{1}}\dots \partial _{i_{k}}\varphi =0$, then
$$
\begin{matrix} 
F^{2}(\wedge (n))\subset F(\varphi \Lambda (n))\subset F(\varphi )
\Lambda (n)+\\
\varphi \sum\limits^{}_{1\le l\le k-1}\; \; \sum\limits ^{}_{(j_{2}, \dots
, j_{l})\subset (i_{1}, \dots, i_{k})}(\partial _{j_{1}}\dots
\partial _{j_{l}}\varphi )\Lambda (n)=0.
\end{matrix} 
$$

Thus, $\str F$ can be nonzero only if $\varphi =\xi _{i_{2}}\dots
\xi _{i_{k}}$.
 
Let us calculate $\str F$ for $F=\xi _{1}\dots \xi _{k}\partial _{1}\dots
\partial _{k}$ and $k<n$. We have
$$
F|\oplus _{0\le i\le k-1}\Lambda ^{i}(n)=0;\quad
\tr F|\Lambda ^{k+i}(n)=\binom{n-k}{i}\;  \text{for}\; 
0\le i\le n-k
$$
implying $(-1)^{k}\str F=\sum^{}_{0\le i\le n-k}(-1)^{i}
\binom{n-k}{i}=0$.
 
2) Follows easily from [BL]. \end{proof}
 
Proof of Theorem 1.3.1. Cases 2) and 3) had been actually considered
even before we have formulated the theorem. In 4) it clearly suffices to prove the
maximality of $\fg $ in $\fgl (V)$. We will prove this together with 1).
 
Let $\fh$ be an intermediate Lie subsuperalgebra: $\fg \subset \fh \subset \fgl
(V)$. Consider two cases: 

$1^{\circ}$. Let $\dim V_{1}=(1, 0)$ or $(0, 1)$, $n>1$, i.e., $V=\Lambda
(n)$  or $\pi (\Lambda (n))$; $\fg =\Lambda (n)\supplus  \fvect (0|n)$ and  $\fgl
(V)\cong \Diff (n)_{L}$. Let $\alpha =(i_{1}, \dots, i_{k})$ be a multiindex,
$D_{\alpha }=\partial _{i_{1}}\dots \partial _{i_{k}}$, $D^{i}_{\alpha
}=
\left\{\begin{matrix}\partial _{i_{1}}\dots \hat \partial _{i}\dots \partial
_{i_{k}}&\text{if $i\in \alpha $}\\ 0&\text{otherwise.}\end{matrix}\right.$
Then
$$
[\sum \varphi _{\alpha }D_{\alpha }, \xi _{i}]=\sum (-1)^{p(D_{\alpha})}
\varphi _{\alpha }D^{i}_{\alpha }\eqno{(1.5)} 
$$

a) It follows from (1.5) that by commuting with functions
$\xi _{1}, \dots, \xi _{n}$ we can obtain an arbitrary 2nd order differential
operator from any differential operator of order $\ge 2$. It
follows that if $\fh \neq \fg$, then $\fh $ contains at least 
one 2nd order operator.
 
b) The representation of $\fk =\fvect (0|n)\subset \fg$ in 
$\Diff_{2}(n)$ is
ireducible if $n>2 $ by Lemma 1.3.2. Therefore, if $\fh \neq \fg$, then 
$\fh$ contains $\Diff_{2}(n)$ if $n>2$ and contains 
$\widetilde{\Diff}_{2}(2)$ if $n=2$.
 
c) Finally, it is not difficult to see that $\Diff_{2}(n)$
generates $\sum^{}_{2\le i\le n-1}\Diff_{i}(n)+\widetilde{\Diff}_{n}(n)$ and,
therefore, if $\fh \neq \fg$, then $\fh =\fsl (V) $.
 
$2^{\circ}$ Let $\dim V_{1}=(r, s)$ and $r+s\ge 2$. a) It follows from (1.4)
that if $A\in \fgl (V_{1})$, $\varphi \in \Lambda (n)$, $D\in \Diff (n)$, then
$$
[A\otimes \varphi, 1\otimes D]=A\otimes [\varphi, D].
$$
Since any operator of the form $A\otimes \varphi$ belongs to the
considered subalgebra $\fg$, it follows that if $\fh \neq 
\fg$, then $\fh $ contains at least one operator with the first component
different from 1.
 
b) Formulas (1.4) and (1.5) imply that commuting with
operators of the form $1\otimes \xi _{i}$ enables one to obtain from any
operator of order $\ge 2$ a 1st order operator. Taking a) into
account we conclude that $\fh$ must contain at least one
1st order operator which does not belong to $\fg$.
 
c) The Lie superalgebra $\fg$ contains as a Lie subalgebra the direct sum of Lie
superalgebras 
$$
\begin{matrix} 
\fgl (V_{1})&\otimes 1\oplus 1\otimes \fvect (0|n) 
&\text{if}&n\neq 1 \\
\fgl (V_{1})&\otimes 1\oplus 1\otimes <\partial >&\text{if}&n=1
\end{matrix} 
$$
and in the space $\fgl (V_{1})\otimes \fvect (0|n)$ the 
tensor product of the adjoint representations of Lie superalgebras $\fgl (V_{1}) $ and
$\fvect (0|n)$ acts. Therefore, $\fh$ contains the set $U_{1}
\otimes U_{2}$, where $U_{1}$ is ideal of $\fgl (V_{1})$ different from
$\Cee\cdot 1$ and $U_{2}$ is an ideal  of $\fvect (0|n)$. Since we have
excluded the case $\dim V_{1}=(1, 1)$,  this means that $U_{1}\supset \fsl
(V_{1})$ and $U_{2}=\fvect (0|n)$ for $n\neq 1$ and $U_{2}\supset \Cee
\partial _{1}$ for $n=1$ (due to simplicity of $\fvect (0|n)$ for $n\neq 1$).
 
Let $A\in \fgl (V_{1})$, $B\in \fsl (V_{1})$. Then taking the bracket 
of $A\otimes \xi$ and $B\otimes \partial$ according to $(1.4)$:
$$
[A\otimes \xi, B\otimes \partial ]=(-1)^{p(B)}([A, B]\otimes \xi \partial +
(-1)^{p(A)p(B)}BA\otimes 1)
$$
we see that $U_{2}$ coincides with $\fvect (0|1)$ for $n=1$, too.
Therefore, $\fh \supset \fsl (V_{1})\otimes \fvect (0|n)$.
 
d) Let  $D, D'\in \fvect (0|n)$. Set 
$$
A_{0}=\left\{\begin{matrix}\diag (1, -1, 0, \dots,
0)\in \fsl (V_{1})&\text{for $r>1 $}\\ 
\diag (0, 1, -1, 0, \dots, 0)\in \fsl (V_{1})&\text{for }
r=1. \end{matrix}\right.
$$  
Then $[A_{0}\otimes D, A_{0}\otimes
D']=A^{2}_{0}\otimes[D, D']$, and $\str A^{2}_{0}\neq 0$. This implies 
$$
\fh \supset \left\{ \begin{matrix} \fgl (V_{1}) \otimes \fvect (0|
 n) &\text{if}\; n>1 \\
\fgl (V_{1}) \otimes <\partial >+\fsl (V_{1})\otimes <\xi \partial > 
&\text{if}\; n=1. \end{matrix} \right.
$$

e) Further, let us carry out the induction on the order
of the differential operator. Let $\fh \supset \sum^{}_{0\le i\le k}
\fgl (V_{1})\otimes \Diff_{i}(n)$. By bracketing $\fgl (V_{1})\otimes \fvect
(0|n)$ according to (1.4)  we get $\fh \supset \fsl (V_{1})\otimes
\Diff_{k+1}(n)$. By bracketing $A_{0}\otimes D_{1}$ with $A_{0}\otimes D_{k+1}$
for $D_{1}\in \fvect (0|n)$ and $D_{k+1}\in \Diff_{k+1}(n) $ we get an operator
$A^{2}_{0}\otimes [D_{1}, D_{k+1}]\in \fgl (V_{1})\otimes \Diff_{k+1}(n)$ that
does not belong to $\fsl (V_{1})\otimes \Diff_{k+1}(n)$. 

Let us consider how $\fgl (V_{1})\otimes 1$ acts in the space $\fgl
(V_{1})\otimes \Diff_{k+1}(n)$. We see that
$$
\begin{matrix} 
\fh &\supset&\fsl (V_{1})\otimes \Diff_{k+1}(n)+
\fgl (V_{1})\otimes [\fvect (0|n), \Diff_{k+1}(n)]= \\
&&\\
&=&\left\{\begin{matrix} 
\fgl (V_{1}) \otimes \Diff_{k+1}(n)& \text{if}& k+1\neq 1 \\
\fgl (V_{1}) \otimes \widetilde{\Diff}_{n}(n)+\fsl (V_{1})
\otimes \Span(\xi _{1}\dots
\xi _{n}\partial _{1}\dots
\partial _{n})& \text{if}& k+1=n.\; \qed \end{matrix}  \right. \end{matrix} 
$$

\ssec{1.4} Let $\fhei (0|2n)$ be generated by odd $\xi _{1}, \dots , \xi _{n}$,
$\eta _{1}, \dots , \eta _{n}$ and an even $z$, see sec. 0.2.

The nonzero irreducible finite dimensional
representation of $\fhei (0|2n)$ that maps $z$ to the operator of
multiplication by 1 is realized in the superspace $\Lambda (n)$ of
functions in $n$ odd variables $\xi =(\xi _{1}, \dots , \xi _{n})$. The
normalizer of $\fhei (0|2n) $ in $\fgl (\Lambda (n))$ is 
 $\fg =\fhei (0|2n)\supplus  \fo (2n)$; it acts in
the spinor representation of $\fo (2n)$, or, in terms of
differential operators: $\fg=\Span (1, \xi, \partial, \xi \partial,
\partial\partial, \xi\xi)$.
 
Let $n=3$. Then $\fg$ is contained in $\tilde\fg =
(\fhei (0|6)\subplus  V)\subplus  \fo (6)$
(sum as $\fo (6) $-modules), where the highest weight of the
$\fo (6)\simeq \fsl (4)$-module $V$ is $(2, 0, 0)$, i.e., 
$\tilde \fg$ is isomorphic to
the nontrivial central extension $\fas$ of $\fspe (4)$. (Observe, that
$\fsl (4) $-module $\fhei (0|6)$ is the direct sum of the trivial
module and the exterior square of the dual to the
standard 4-dimensional $\fsl(4)$-module.)
 
\begin{Theorem} $\fg $ is a maximal
subalgebra in $\fsl(\Lambda (n))$ for $n\ge 2 $ and $n\neq 3$. If $n=3$, then
$\fg \subset  \tilde \fg\cong\fas$, and $\fas$ is maximal in $\fsl (\Lambda (n))$.
\end{Theorem}

\begin{proof} Let us realize again $\fsl (\Lambda (n))$ by differential
operators
$$
\fsl (\Lambda (n))=\oplus _{i\le n-1}\Diff_{i}(n)\oplus 
\widetilde{\Diff}_{n}(n).\eqno{(1.6)}
$$

Let $\fh$ be a Lie subsuperalgebra in $\fsl (\Lambda (n))$ 
containing $\fg$ and strictly greater than it.
 
$1\degree$ First, let us show that $\fh $ contains an element $h\not\in \fg$
whose highest term with respect to the grading (1.6) is of
degree 1. Let $h\in \fh $ and $h\not\in \fg$. Note that
$$
[\varphi \partial _{i_{1}}\dots \partial _{i_{k}}, \xi
_{i}]=\left\{\begin{matrix}  0 &
\text{for}&i\not\in (i_{1},  \dots, i_{k}) \\ 
\pm \varphi \partial _{i_{1}}\dots \hat \partial _{i}\dots \partial _{i_{k}} &
\text{for}& i\in (i_{1}, \dots , i_{k}) \end{matrix}  \right.
$$
 
Since $\xi _{i}\in \fg$ for all $i=1, \dots
, n$, it follows that if $\deg h\ge 2$, then
commuting with $\xi _{i}$ we may obtain an element $\tilde h$ of degree 1
from $h$. Now, since $\xi _{i}\xi _{j}\in \fg $ for all $(i, j)\in (1, 
\dots, n)$ and
$$
\begin{matrix} [\varphi \partial _{i_{1}}\dots
\partial _{i_{k}}, \xi _{i}\xi _{j}]\\
= \left\{ 
\begin{matrix}  0 & \text{for}& i, j
\not\in (i_{1}, \dots, i_{k}) \\
\pm \xi _{j}\varphi \partial _{i_{1}}\dots\hat \partial _{i}\dots
\partial _{i_{k}} & \text{for}& i\in (i_{1}, \dots
, i_{k}), j\not\in (i_{1}, \dots, i_{k}) \\
\pm \xi _{i}\varphi \partial _{i_{1}}\dots\hat \partial _{j}\dots
\partial _{i_{k}} & \text{for}& 
i\not\in (i_{1}, \dots, i_{k}), j\in (i_{1}, \dots, i_{k}) \\
\pm \xi _{i}\varphi \partial _{i_{1}}\dots\hat \partial _{j}\dots
\partial _{i_{k}}&&\\
\pm \xi _{j}\varphi \partial _{i_{1}}\dots
\hat \partial _{i}\dots\partial _{i_{k}}\pm \varphi \partial _{i_{1}}\dots
\hat \partial _{i}\dots\hat \partial _{j}\dots\partial _{i_{k}} 
& \text{for}& i, j\in (i_{1}, \dots, i_{k}), \end{matrix}  \right.
\end{matrix} 
$$
we may assume that $\tilde h\not\in \fg$. If $\deg h=0$, i.e., $h\in 
\Lambda (n)$, it suffices to consider brackets of the form
$$
[\partial _{i}\partial _{j}, h]=\pm \partial _{i}(h)\partial _{j}\pm 
\partial _{j}(h)\partial _{i}\pm \partial _{i}\partial _{j}(h)
$$
and notice that since $h\not\in \fg$, then $\deg h\ge 3 $ with respect 
to the grading in $\Lambda (n)$.

\vskip 0.2 cm
 
$2\degree$ Let $n=2$. We have got that $\fh$ contains an element
of the form $\xi _{1}\cdot \xi _{2}(\alpha \partial _{1 }+\beta \partial
_{2})$  for some $(\alpha, \beta )\neq (0, 0)$. Since $\fgl (2)=<\xi
_{i}\partial _{j}>$ acts irreducibly in the space $<\xi _{1} \xi _{2}\partial
_{1}, \xi _{1}\xi _{2}\partial _{2}>$, then
$\fh \supset \Lambda (2)\oplus \fvect (0|2)$ and by Theorem 1.3.1 
$\fh =\fsl (\Lambda (2))$.

\vskip 0.2 cm
 
$3\degree$. Let $n\ge 3$. By $1^{0}$, we may assume that $\fg$ contains a
set of elements whose highest terms run over $\fsvect (0|n)\subset \fvect(0|n)$.
Acting by them on functions contained in $\fg$ we see that $\fh$
contains $T^{0}(n)\oplus \fsvect (0|n)$, where 
$T^{0}(n)\cong\Lambda (n)/<\xi _{1}\dots\xi _{n}>$ is the superspace of
functions with integral 0.
 
For $n=3$ it is easy to see that $\fg \oplus T^{0}(3)\oplus \fsvect 
(0|3)$ generates $\tilde\fg$.
 
If $n>3$, then the relation
$$
[\partial _{1}\partial _{2}, \xi _{1}\xi _{2}\xi _{3}]=(\xi _{1}\partial _{1}+
\xi _{2}\partial _{2}-1)\xi _{3}\not\in \fsvect (0|n)
$$
implies that the sets $T^{0}(n)$ and $<\partial _{i}\partial _{j}$, $1\le j\le
n>$ generate $\fvect (0|n)$; hence, $\fh \supset \Lambda (n)\oplus \fvect
(0|n)$ and, by Theorem 1.3.1, $\fh =\fsl (\Lambda (n)) $.

The maximality of $\tilde \fg$ for $n=3$ is obvious. \end{proof}

\section*{\protect\S 2. Proof of Main Theorem}  

In this section $\fg$ is an irreducible matrix Lie superalgebra, $\rho$ its
standard representation in superspace $W$. Let $\fk$ be an ideal of $\fg$. We
will assume that $\dim W>1$ (because the case $\dim W=1$ is trivial). Our proof
of Main Theorem is largely based on the following Theorem 2.0 --- the
superization of a statement well-known to the reader from Dixmier's book [Di].

\ssbegin{2.0}{Theorem} Let $\fg$ be an irreducible matrix Lie
superalgebra, $\rho$ its standard representation (in particular,
$\rho$ is faithful). Let $\fk $ be an ideal of $\fg $ and $\dim
\fk>1$. Then $3$ cases are possible:

{\em A)} $\rho |\fk $ is irreducible;

{\em  B)} $\rho |\fk $ is a multiple of an irreducible $\fk
$-module $\tau $ and the multiplicity of $\tau $ is $>1$;

{\em C)} there exists a subalgebra $\fh\subset\fg$ such that $\fk\subset
\fh$ and $\rho\equiv \ind^{\fg }_{\fh }\sigma $ for an irreducible
$\fk$-module $\sigma $ .
\end{Theorem}

We will say that the {\it representation $\rho$ is of type $A$, $B$, or
$C$ with respect to the ideal} $\fk$ if the corresponding case holds. Lemmas
2.4, 2.1 and 2.2 deal with types A, B and C, respectively.

Proof largely follows the lines of [Di] with one novel case: irreducible modules
of $Q$-type might occure. Fortunately, their treatment is rather straightforward
and I would rather save paper by ommiting the verification. 

\ssbegin{2.1}{Lemma} {\em {\bf B)} 1)} If $\rho$ is of type B with respect to
the ideal $\fk$, then either $\fg\subset \fgl(V_1)\bigodot\fgl(V_2)$ or
$\fg\subset
\fq(V_1)\bigodot\fq(V_2)$ for some $V_1$ and $V_2$. 

{\em 2)} If, moreover, $\fg\subset\fq(V)$, then $\fg\subset
\fq(V_1)\bigodot\fgl(V_2)$ for some $V_1$, $V_2$;

{\em 3)} If $\omega$ is a nondegenerate $2$-form on $V$ and $\fg\subset
\faut(\omega)$, then $\fg\subset \faut(\omega_1)\oplus \faut(\omega_2)$, where
$\omega=\omega_1\otimes \omega_2$
\end{Lemma}

\begin{proof} 1) Let $V$ be the superspace of representation $\tau$. Then
$V=V_1\otimes U$ for some $U$ and 
$$
\rho(h)(v\otimes u)=\tau(h)v\otimes u\text{ for any}\ h\in\fk, \; v\in V_1, 
\; u\in U
$$
i.e., 
$$
\text{the operators $\rho (h)$ are of the form $\tau(h)\otimes 1$.}\eqno{(2.1)}
$$ 

On the other hand, $\fg$ is the stabilizer of $\tau$. So for any
$g\in\fg$ there exists $s(g)\in \End V_1$ such that $\tau ([g, h])=[s(g),
\tau(h)]$ for any
$h\in\fk$. 

Let us consider the operator $T(g)=\rho(g)-s(g)\otimes 1$ in $V$. For
$h\in\fk$ we have
$$
[T(g), \rho(h)]=[\rho(g)-s(g)\otimes 1, \rho(h)]=\rho([g, h])-[s(g), \tau(h)]
\otimes 1=0\ , 
$$
i.e., $T(g)$ commutes with all the operators of the form $\rho(h)$, $h\in \fh$
and since $\tau$ is irreducible, $T(g)$ is of the form
$$
T(g)=\left\{\begin{array}{l}
1\otimes A(g)\ \text{ if}\ \tau\text{  is of $G$-type}\\
1\otimes A(g)+J\otimes B(g)\text{  if $\tau$ is of $Q$-type}\ , 
\end{array}\right.
$$
where $J$ is an odd generator of $C(\tau)$, the centralizer of $\tau$.

Therefore, $\rho(g)$ is either of the form
$$
\rho(g)=s(g)\otimes 1+1\otimes A(g), \text{  if $\tau$ is of
$G$-type}\eqno{(2.2)} 
$$ or
$$
\rho(g)=s(g)\otimes 1+1\otimes A(g)+J\otimes B(g)\text{  if $\tau$ is of
$Q$-type}\ .\eqno{(2.3)} 
$$
Formulas $(2.1)-(2.3)$ show that if $\tau$ is of $G$-type, then $\fg$ is
contained in $\fgl(V_1)\bigodot\fgl(V_2)\subset\fgl(V)$, where $V_2=U$, and if
$\tau$ is of
$Q$-type, then $\fg$ is contained in
$\fq(V_1)\bigodot\fq(V_2)\subset\fgl(V)$, where $V_2=\Span(1, J)\otimes U$. 

2) To prove this heading it suffices to notice that if $\tau$ is of type $G$ and
the operators $T(g)=1\otimes A(g)$ determine a $G$-irreducible action on $U$,
then $\rho$ is also of type $G$. Therefore, either
$\fg\subset\fq(V_1)\bigodot\fgl(V_2)$, where $V_2=U$, or
$\fg\subset\fgl(V_1)\bigodot\fq(U)$.

3) To prove this heading, observe that if we fix $u_1, u_2\in U$, then on $V$
there arises a family of $\fk$-invariant 2-forms $\omega_1^{u_{1}, u_{2}}$. Since
$\tau$ is irreducible, each form $\omega_1^{u_{1}, u_{2}}$ is either
nondegenerate or identically zero. Since $\omega$ is nondegenerate, there is a
nondegenerate form $\omega_1=\omega_1^{u_{1}, u_{2}}$ among the lot. In this case
the whole space of forms is 1-dimensional and $\tau$ is a
$G$-irreducible representation. For Lie algebras this is a well-known fact; for
Lie superalgebras we prove it in Appendix 2.

Therefore, $\omega (v_1\otimes u_1, v_2\otimes u_2)=\omega_1(v_1, v_2)\cdot
\omega_2(u_1, u_2)$, where $\omega_2$ is the nondegenerate bilinear form on
$U$. The action of an element $\rho(g)$ of the form (2.2) on $\omega$ gives:
$$
\begin{matrix}
0=(s(g)\otimes 1+1\otimes A(g))\omega(v_1\otimes u_1, v_2\otimes u_2)=\\
\left (\omega_1(s(g)(v_1), v_2)+(-1)^{p(g)p(v_{1})}\omega_1(v_1,
s(g)(v_2))\right)\omega_2(u_1, u_2)=\\
\omega_1(v_1, v_2)\left (\omega_2(A(g)(u_1), u_2)+(-1)^{p(g)p(u_{1})}\omega_2(u_1,
A(g)(u_2))\right).
\end{matrix}
$$
After simplification we get:
$$
s(g)(\omega_1)=c(g)\omega_1;\quad A(g)\omega_2=-c(g)\omega_2,
$$
where $c(g)\in\Cee$. 

Therefore, 
$$
\rho(g)=s(g)\otimes 1+1\otimes A(g)=(s(g)-\frac 12c(g)\cdot\id)\otimes 1+
1\otimes(A(g)+\frac 12c(g)\cdot \id),
$$
where $s(g)-\frac 12c(g)\cdot\id\in\faut(\omega_1)$ and 
$\id\otimes(A(g)+\frac 12c(g)\cdot \id)\in\faut(\omega_2)$.
\end{proof}

\ssbegin{2.2}{Lemma} {\em {\bf C)} 1)} If $\rho$ is of type C with respect to
the ideal $\fk$, then $\fg\subset \fgl(V)\otimes \Lambda
(n)\supplus\fvect(0|n)$ for some $V$ and $n$.  

{\em 2)} If, moreover, $\fg\subset\fq(n)$, then either $\fg\subset
\fq(V)\otimes\Lambda (n)\supplus\fvect(0|n)$ or 

$n=1$ and $\fg\subset
\fgl(V)\otimes\id_{\Lambda(1)}\supplus\Cee(\xi+\partial\xi)$;

{\em 3)} If $\rho$ is of type C with respect to the ideal $\fk$ and $\fg\subset
\faut(\omega)$, then $\fg\subset\faut(\omega_1)\otimes \Lambda(n)\oplus
T^{1/2}(\fvect(0|n))$, where $\omega=\omega_1\otimes \omega_2$ and 
the representation $T^{1/2}$ of $\fvect(0|n)$ (in the superspace of halfdensities)
is given by the formula $T^{1/2}(D)=1\otimes (D+\frac 12\Div D)$ and the form
$\omega_2$ on $\Lambda (\xi)$ is given by the formula
$$
\omega_2(f\sqrt{\vvol(\xi)}, g\sqrt{\vvol(\xi)})=\int fg\cdot \vvol(\xi).
$$
\end{Lemma}

\begin{Remark} 1)
$\fgl(V)\otimes\id_{\Lambda(1)}\supplus\Cee(\xi+\partial\xi)\cong\fgl(V)\bigodot
\fq(1)$. 

2) $\faut(\omega_1)\otimes 1\oplus
T^{1/2}(\fvect(0|n))\subset\faut(\omega_1)\bigodot\faut(\omega_2)$
\end{Remark}

Lemma C follows from the definition of the induced representaiton.

\ssec{2.3} To complete the
proof of Main Theorem it suffices to consider the case when
$\rho$ possesses the following property: for any ideal $\fk\subset \fg$ either
$\rho$ is of type A with respect to $\fk$ or $\rho\vert _\fk$ is the
multiple of a character. 

Let $\fr$ be the radical of the matrix Lie
superalgebras $\fg$.

\begin{Lemma} If $\fr$ is commutative and $\rho$ belongs to type A or
B with respect to $\fr$, then $\dim\fr=0$ or $1$. 
\end{Lemma}

Proof follows from the fact that any irreducible
representation of a commutative Lie superalgebra is 1-dimensional (up to
the change of parity) and the faithfulness of $\rho$. \qed

\ssec{2.4} In the case when
$\fr$ is not commutative, consider the derived series of $\fr$:
$$
\fr\supset\fr_1\supset\dots\supset\fr_k\supset\fr_{k+1}=0,
$$
where $\fr_{i+1}=[\fr_i, \fr_i]$. Clearly, each $\fr_i$ is an ideal in $\fg$
and the last ideal, $\fr_k$ is commutative. 

\begin{Lemma} {\em {\bf A)} 1)} If $\rho$ is of type A with respect to
$\fr_{k-1}$ and $\rho| _{\fr_{k}}$ is scalar, then either
$\fr_{k-1}\cong\fhei(0|2n)$ or $\fr_{k-1}\cong\fhei(0|2n-1)$ and
$V\cong\Lambda(n)$ or $\Pi(\Lambda(n))$ and $\fg\subset
\fhei(0|2n)\supplus\fo(2n)$. 

{\em 2)} If additionally $\fg\subset \fq(V)$, then either
$\tau_{k-1}\cong\fhei(0|2n-1)$ or $\fg\subset\fhei(0|2n-1)\supplus\fo(2n-2)\cong
(\fhei(0|2n-2)\supplus\fo(2n-2))\bigodot \fq(1)$.

{\em 3)} In this case $\fg$ does not preserve any nondegenerate 2-form on $V$.
\end{Lemma}

\begin{proof} Headings 2 and 3 follow from heading 1. Let us prove it.

We see that
if the restriction
$\rho|_{\fr_{i}}$ is irreducible for all $i$, then

1) $\dim\fr_k=1$ because $\rho$ is faithful;

2) $\fr_{k-1}=\fhei (0|m)$ for any $m$;

3) $\rho|_{\fr_{k-1}}$ is irreducible and faithful, so it can be
realized in the superspace of functions, $\Lambda (2n)$, or in
$\Pi(\Lambda (2n))$, where $n=[\frac m2]+1$;

4) $\fg$ is contained in the normalizer of $\fhei (0|2n)$ in its spinor
representation, i.e., 
$\fg\subset\fhei (0|2n)\subplus  \fo(2n)$. \end{proof}

\ssbegin{2.5}{Lemma} Let $\dim\fr\leq 1$, i.e., $\fg$
is either semisimple or a nontrivial central extension of a semisimple Lie
superalgebra but NOT an almost simple or a central extension of an almost
simple Lie superalgebra. Then we can always choose an ideal $\fk$ such that  
$\rho$ is of type B or C with respect to $\fk$.
\end{Lemma}

We proove this Lemma in Appendix 1.

Lemmas 2.1 (Lemma B), 2.2 (Lemma C), 2.3, 2.4 (Lemma A) and Lemma 2.5 put
together prove Main Theorem. \qed

\section*{\protect\S 3. Irreducible non-simple maximal matrix subalgebras 
of $\fq(n)$}

\ssec{ 3.1} Let $\dim W=(n, n)$, $n\ge 2$ and
$J=\begin{pmatrix} 0& 1_{n}\\ -1_{n}& 0\end{pmatrix}$
(so, $J^{2}=-1_{V}$ and $p(J)=\overline 1)$. Consider the centralizer
$C(J)=\{A\in \fgl (V): [A, J]=0\}$ of $J$. Notice that on the set $C(J)$, 
as well as on the set of the all operators, two
structures can be introduced: that of the associative
superalgebra $Q(W)$ and that of the Lie superalgebra
$\fq (W)=Q(W)_L$.

Now, let $V=V_{1}\otimes V_{2}$ and $\dim V_{1}=(n, n)$; let $\dim V_{2}=(n_2,
m_2)$ be arbitrary. Given an odd operator $J_{1}\in \End (V_{1})$ such that
$J^{2}_{1}=-1_{V_{1}}$
(and, therefore, a superalgebra $C(J_{1})$ being defined), then
the operator $J=J_{1}\otimes 1$ in $V$ is defined such that $J^{2}=-1_{V}$ and
$p(J)=\overline 1$. Observe that
$$
Q\Mat (V)\cong Q \Mat (V_{1})\bigodot\Mat (V_{2}).\eqno{(3.1)}
$$
Consider the Lie superalgebra $\fg =\fq (V_{1})\bigodot \fgl (V_{2})$. 
Clearly, $\fg \subset \fq (V)$. 

\ssbegin{3.1.1}{Lemma} The Lie superalgebra $\fg$ is irreducible on
$V$ if $n_2\neq m_2$.
\end{Lemma}

\begin{proof} The operators of the form $1\otimes B$ for$B\in \fgl (V_{2})$ 
commute with the operators of $\fgl (V)$ that are of the form $C\otimes 1$ for 
$C\in \fgl (V_{1})$ and only with them, while operators of the form
$A\otimes 1$ for $A\in \fq (V_{1})$ commute with the operators of the form
$1\otimes C$ and $J_{1}\otimes C$, where $C\in \fgl (V_{2})$. Therefore, the
algebra of intertwining operators of the considered representation of $\fg$ is
generated by two operators: 1 and $J_{1}\otimes 1$. \end{proof} 

\ssbegin{3.1.2}{Theorem} The Lie subsuperalgebra $\fg
=\fq(V_{1})\bigodot
\fgl  (V_{2})$ is maximal in $\fq (V_{1}\otimes V_{2})$.
\end{Theorem}

\begin{proof} Formula (3.1) enables to actually repeat the proof of Theorem
1.1.

We must only notice that like $\fgl (W)$,  the Lie superalgebra $\fq(W)$ 
contains two ideals only: that of \lq\lq odd constants" and $\fsq(W)=\{X\in \fq
(W)|\qtr X=0\}$ which is not endowed with the natural associative
algebra structure. Besides, if $g\in \fq (V_{1})$ but $g\not\in \fsq (V_{1})$,
then $g\otimes 1\not\in \fsq (V)$. 
\end{proof} 

\ssec{3.2} Let $V_1$ be an $(m, m)$-dimensional linear superspace and
$\fq(V_1)=\fq(m)=C(J)$ be the queer Lie superalgebra with the standard
representation in $V_1$. Let $V=V_1\otimes\Lambda (n)$.

\begin{th*}{ Theorem} The Lie subsuperalgebra $\fg =\fq(V_{1})\otimes\Lambda
(n)\subplus  \fvect(0|n)$ is maximal in $\fq (V)=C(J\otimes 1)$.
\end{th*}

\begin{proof} We can repeat the proof of Theorem 1.3.1 by replacing $\fgl(V_1)$
with $\fq(V_1)$ and $\fsl(V_1)$ with $\fsq(V_1)$ provided in step d) we 
consider  
$$
[\Pi A_0\otimes D,\quad  A_0\otimes D']=\Pi A_0^2[D, D']
$$
instead of $[A_0\otimes D, A_0\otimes D']$. 
\end{proof}

\ssec{3.3} Adding together Theorems 3.1.2, 3.2 and heading $2\degree$
of Main Theorem we get the following statement.

\begin{Theorem} If $\fg
\subset\fq(n)=\fq(V)$ is a maximal in
$\fq(n)$ irreducible Lie superalgebra, not almost simple nor a central extension
of an almost simple, then either
$\fg=\fq(V_1)\bigodot\fgl(V_2)$ for some $V_1$ and $V_2$ such that $V=V_1\otimes
V_2$ and $\dim V_2>1$ or

$\fg =\fq(V_{0})\otimes\Lambda
(n)\subplus  \fvect(0|n)$ for some $V_0$ and $n$ such that $V=V_0\otimes
\Lambda (n)$.
\end{Theorem}

\section*{\S 4. Lie superalgebras that preserve a bilinear form}

Let a nondegenerate homogeneous supersymmetric bilinear
form $\omega $ be given in a superspace $V$. Sometimes, in a fixed basis,  we
will represent
$\omega$ by its matrix $F$. Consider two objects associatted with $\omega $:

1) The space $\faut (\omega )=\faut (F)$ of operators preserving
$\omega $:
$$
\faut (\omega )=\{A\in \fgl (V)\mid \omega (Ax, y)+(-1)^{p(A)(p(x)+
p(\omega ))}\omega (x, A^{st}y)=0\}
$$
or, in the matrix form, 
$$
\faut(F)=\{A\in \fgl (V)\mid AF+(-1)^{p(A)p(F)}FA^{st}=0\}.
\eqno{(4.1)}
$$

2) The space $\symm (\omega )=\symm (F)$ of operators symmetric
with respect to $\omega $:
$$
\symm (\omega )=\{A\in \fgl (V)\mid \omega (Ax, y)=
(-1)^{p(A)(p(x)+p(\omega ))}\omega (x, A^{st}y)\}
$$
or, in the matrix form, 
$$
\symm (F)=\{A\in \fgl (V)\mid AF=(-1)^{p(A)p(F)}FA^{st}\}.
\eqno{(4.2)}
$$
Direct calculations yield the following statement:

\ssbegin{4.1}{Lemma} {\em 1)} The space $\faut (\omega )$ is endowed with a
natural Lie superalgebra structure while the superspace $\symm (\omega )$ is
endowed with the stucture of an $\faut (\omega )$-module. Moreover, $[\symm 
(\omega ), \symm (\omega )]\subset \faut (\omega )$. 

$2)$ Set $\{\cdot, \cdot \}: A, B\tto AB+(-1)^{p(A)p(B)}BA$. Then
$$
\{\faut (\omega ), \faut (\omega )\},\quad \{\symm (\omega ), 
\symm (\omega )\}\subset \symm (\omega ); \quad
\{\faut (\omega ), \symm (\omega )\}\subset \faut (\omega );
$$
$$
\begin{matrix} 
\faut (\omega _{1}\otimes \omega _{2})&=
\faut (\omega _{1})\otimes \symm (\omega _{2})+\symm 
(\omega _{2})\otimes \faut (\omega _{2}); \\
\symm (\omega _{1}\otimes \omega _{2})&=
\faut (\omega _{1})\otimes \faut (\omega _{2})+\symm 
(\omega _{1})\otimes \symm (\omega _{2}). \end{matrix}  
\eqno{(4.3)}
$$
$3)$ Any $A\in \fgl (V)$ can be represented as the sum $A=A^{a}+A^{s}$, 
where $A^{a}\in \faut (\omega )$, $A^{s}\in \symm (\omega )$.
\end{Lemma}

Set
$$
\fs\faut (\omega )=\faut (\omega )\cap \fsl (V),\quad\quad \ssym (\omega )
=\symm (\omega )\cap \fsl (V).
$$
Notice that if $p(\omega )=\overline 0$, then $\faut (\omega )=
\fs\faut (\omega )$, and if $p(\omega )=\overline 1$,
then $\symm (\omega )=\ssym (\omega )$.

\begin{th*}{Theorem} Let Lie superalgebras $\fs\faut (\omega _{1})$ and 
$\fs\faut(\omega _{2})$ be simple. Then Lie subalgebra $\fg =\faut (\omega
_{1})\oplus \faut  (\omega _{2})$ is maximal in $\faut (\omega _{1}\otimes \omega
_{2})$ if $p(\omega _{1})+p(\omega _{2}) =\overline 0$ or if $p(\omega
_{i})=\overline 0$, $p(\omega _{j})=\overline 1$ and  $\dim(V_{i})_{\bar 0}\neq
\dim (V_{i})_{\bar 1}$ for $i, j=1, 2$. If  $p(\omega _{i})=\overline 0$,
$p(\omega _{j})=\overline 1$ and $\dim (V_{i})_{\bar 0}=\dim (V_{i})_{\bar 1}$,
then $\fg
$ is maximal in  $\fs\faut (\omega _{1}\otimes \omega _{2})$.
\end{th*}

For exact answer in the standard format see table from Th. 0.2.4.

\begin{proof} Let $\ff $ be a Lie superalgebra such that
$\fg \subset \ff \subset \faut (\omega _{1}\otimes \omega _{2})$ 
and $h\in \ff $ but $h\not\in \fg $. It follows from (4.3) that
$$
h=\sum A^{a}_{i}\otimes B^{s}_{i}+\sum A^{s}_{j}\otimes B^{a}_{j}
$$
for some $A^{a}_{i}\in \faut (\omega _{1})$, $B^{s}_{i}\in \symm
(\omega _{2})$, $A^{s}_{j}\in \symm (\omega _{1})$, $B^{a}_{j}\in 
\faut (\omega _{2})$. Let us bracket $h$ with elements of $\fg$:
$$
\begin{matrix} 
[C\otimes 1, h]&=&\sum [C, A^{a}_{i}]\otimes B^{s}_{i}+\sum [C,
A^{s}_{j}]\otimes  B^{a}_{j} \\
{}[1\otimes D, h]&=&\sum (-1)^{p(D)p(A^{a}_{i})}A^{a}_{i}\otimes [D, B^{s}_{i}]+
\sum (-1)^{p(D)p(A^{s}_{j})}A^{s}_{j}\otimes [D, B^{a}_{j}]. \end{matrix} 
$$

Since $\fs\faut (\omega _{1})$ and $\fs\faut (\omega _{2})$ are simple and 
(as follows from [K]) the sole $\faut (\omega )$-invariant superspaces in
$\symm (\omega )$ are the space of contants and $\ssym (\omega )$,
it follows that
$$
\ff \supset (\fs\faut (\omega _{1})\otimes \ssym (\omega _{2}))\oplus 
(\ssym (\omega _{1})\otimes \fs\faut (\omega _{2}))=\fa\oplus 
\fb.
$$
Now notice that $\fsl (V)$ is not closed with respect to
$\{\cdot , \cdot\}$.

Let $p(\omega _{1})=p(\omega _{2})=\overline 0$. Then $\faut (\omega _{i})=
\fs\faut (\omega _{i})$ for $i=1, 2$. Let
$A\otimes B, C\otimes D\in \fa$. We have:
$$
[A, C]\otimes \{B, D\}\in \faut (\omega _{1})\otimes\symm  (\omega
_{2}),\quad \quad \{A, C\}\otimes [B, D]\in\symm (\omega _{1})\otimes
\faut(\omega _{2}). 
$$
Since $\fsl (V)$ is not closed with respect to
$\{\cdot , \cdot\}$ the above formulas imply that
$$
\ff \supset \faut (\omega _{1})\otimes \symm (\omega _{2})\oplus 
\symm (\omega _{1})\otimes \faut (\omega _{2})=\faut (\omega _{1}
\otimes \omega _{2}).
$$
Let $p(\omega _{1})=p(\omega _{2})=\overline 1$ and, therefore, $\ssym 
(\omega _{i})=\symm (\omega _{i})$. Let
$A\otimes B\in \fa , C\otimes D\in \fb$. Then
$$
[A, C]\otimes \{B, D\}\in \symm (\omega _{1})\otimes \faut 
(\omega _{2}),\quad \quad  \{A, C\}\otimes [B, D]\in \faut (\omega _{1})\otimes 
\symm (\omega _{2})
$$
which means that $\ff=\faut (\omega _{1}\otimes \omega _{2})$.

Finally, let $p(\omega _{1})=\overline 0, p(\omega _{2})=\overline 1$. Then
$\fs\faut (\omega _{1})\otimes \ssym (\omega _{2})=\faut (\omega
_{1}) \otimes \symm (\omega _{2})$. Taking the bracket of $A\otimes B\in
\fa$ with $C\otimes D\in \fb$, we see that 
$$
\ff \supset \faut (\omega
_{1})\otimes \symm (\omega _{2})\oplus \ssym (\omega _{1}) \otimes
\faut (\omega _{2})
$$
and taking the bracket of $A\otimes B, C\otimes D\in
\fb$ we get 
$$
\ff \supset \faut(\omega _{1})\otimes \symm (\omega _{2})+\symm 
(\omega _{1})\otimes \fs\faut (\omega _{2}).
$$
These two inclusions together mean that $\ff \supset \fs\faut (\omega _{1}
\otimes \omega _{2})$. This completes the proof when $\dim (V_{1})_{\bar 0}=
\dim (V_{1})_{\bar 1}$, while for $\dim (V_{1})_{\bar 0}\neq \dim (V_{1})_{\bar
1}$ it suffices to observe that  $\fg$ is not contained in $\fsl(V)$. 
\end{proof}

\section*{Appendix 1: Proof of Lemma 2.5}

Let $\dim\fr=0$, i.e., $\fg$
is semisimple. As $\fg$ is not almost simple, then due to sec. A2.6 the
alternative arises: either $\fg$ contains an ideal $\fk$  of the form
$$
\fk=\fh\otimes \Lambda (n)\quad\text{with simple $\fh$ and $n>0$}\eqno{(A1.1)}
$$
or $\fg=\mathop{\oplus}\limits_{j\leq k}\fs_j$, where each
$\fs_j$ is almost simple and $k>1$.

\ssbegin{A1.1}{Lemma} If $\fg=\mathop{\oplus}\limits_{j\leq k}\fs_j$ and $k\geq 2$,
then any irreducible faithful representation of $\fg$ is of type B with respect to
any its ideal
$\fs_j$.\end{Lemma}

\begin{proof} Since the stabilizer of any irreducible representation of
$\fs_j$ is the whole $\fg$, the type of any irreducible representation of
$\fg$ with respect to $\fs_j$ can be either A or B, see sec. 2.0. Due to
faithfulness case A is ruled out.
\end{proof}

For $\fk$ take $\fa\fsi\otimes\Lambda(n)$.

\ssbegin{A1.2}{Lemma} Let $\fh$ be a simple Lie superalgerba and
$\fk=\fh\otimes\Lambda(n)$. Then $\fk$ has no faithful irreducible finite
dimensional representations.
\end{Lemma}

For proof see A1.4.

\ssbegin{A1.2.1}{Corollary} If $\fg$ contains an ideal $\fk$ of the form
$(A1.1)$, then $\fg$ can not have any faithful irreducible finite dimensional
representation of type A with respect to the ideal $\fk$.
\end{Corollary}

\ssbegin{A1.2.2}{Corollary} Lemma A$2.1$ and Corollary A$2.2.1$ prove Lemma $2.5$ for
semisimple Lie superalgebras.
\end{Corollary}

\ssbegin{A1.3}{Lemma} If $\fh$ is a simple Lie superalgebra,
then
$$
[\fh_{\od}, \fh_{\od}]=\fh_{\ev}\quad
{\it and}\quad [\fh_{\ev}, \fh_{\od}]=\fh_{\od}\ .
$$
\end{Lemma}

Proof is left to the reader. ({\it Hint}: show that
$[\fh_{\od}, \fh _{\od}]\oplus\fh_{\od}$ and
$\fh_{\ev}\oplus [\fh _{\ev}, \fh_{\od}]$ are ideals in
$\fh$.) \qed 

\ssec{A1.4. Proof of Lemma A1.2}
For $n> 0$ the Lie superalgerba $\fg$ contains
supercommutative ideal
$$
\fn_1=\fh\otimes\xi_1\dots\xi_n.
$$
Moreover, if $n>2$, then $\fg$ additionally contains a
supercommutative ideal
$$
\fn_2=\fh\otimes\Lambda^{n-1}(\xi)\oplus \fn_1 .
$$
Thanks to Theorem 2.0, any irreducible representation $\rho$ of $\fk$
is obtained as follows: take an irreducible representation (character $\chi$) of
an ideal
$\fn_i$, the subsuperalgebra $\fst _{\fk}(\chi)$, an irreducible
representation $\tilde{\rho}$ of $\fst_{\fk} (\chi)$ whose restriction
onto $\fn_i$ is a multiple of $\chi$. Then
$\rho=ind^{\fk}_{\fst_{\fk}(\chi)} (\tilde{\rho})$ for some $\rho$.

Let us show that if $\dim\rho<\infty$ then
$\chi|_{\fh\otimes\xi_1\cdots \xi_n}=0$. Since $\fh\otimes\xi_1\cdots
\xi_n$ is an ideal in $\fg$, this would contradict the faithfulness of
$\rho$. Indeed, $\dim\rho<\infty$ if and only if 
$\fst_{\fk}(\chi)\supset\fk_{ \ev }$.

1) $n=2k+1$. For any $f\in \fh$ and $g=h\otimes 1\in
\fh_0\otimes 1$ we have $0=\chi ([g, f\otimes\xi_1\cdots
\xi_n])=\chi ([h, f]\otimes\xi_1\cdots \xi_n)$ implying that
$$
\chi |_{[\fh_{\ev}, \fh_{\od}]\otimes\xi_1\cdots \xi_n}\overset{\text{ (Lemma
A1.3)} }{=}
\chi|_{(\fh_{\od}\otimes\xi_1\cdots \xi_n)}=0.
$$
Since $\chi$ is a character, then $\chi|_{\fh_{\ev}\otimes\xi_1\cdots
\xi_n}=0$, i.e.,
$\chi|_{\fh\otimes\xi_1\cdots \xi_n}=0$.

2) $n=2k>2$. For any
$$
\begin{matrix}
g=h\otimes l_1(\xi)\in \fh_{\od}\otimes
\Lambda^1(\xi)\subset\fg _{\ev}\hspace{1.5cm}\\
g_1=h_1\otimes l_{n-1}(\xi)\in \fh_{\od}\otimes\Lambda^{n-1}(\xi)\subset
(\fn_2)_{\ev}
\end{matrix}
$$
we have
$$
0=\chi([g, g_1])=\pm \chi([h, h_1])\otimes l_1(\xi)l_{n-1}(\xi),
$$
i.e.,
$$
\chi|_{[\fh_{\od}, \fh_{\od}]\otimes\xi_1\cdots \xi_n}=\chi |
_{\fh_{\ev}\otimes \xi_1\cdots \xi_n}=0\ .
$$
But $\chi$ is a character; therefore,
$$
\chi|_{\fh_{\od}\otimes\xi_1\cdots \xi_n}=0
$$
and, finally,
$$
\chi|_{\fh\otimes\xi_1\cdots \xi_n=0}\ . \qed
$$

3) $n=2$. Then $\fk$ contains a commutative ideal
$\fn=\fh\otimes\xi_1\xi_2;$ let $\chi$ be a character of $\fn$.

If $\chi$ generates a finite dimensional representation of $\fg$ then
$\fst_{\fg(\chi)}\supset\fg_{\ev}$, i.e.,
$$
\chi([f, g\otimes\xi_1\xi_2])=0\; \text{ for any}\; f\in \fh_{\ev}, g\in
\fh . 
$$

If $g\in \fh_{\od}$, then formula $(A1.1)$ holds automatically by
the parity considerations. Let $g\in \fh_{\ev}$. Then,
nevertheless,
$$
\chi([f, g\otimes\xi_1\xi_2])=\chi([f, g]\otimes\xi_1\xi_2)=0
$$
and, therefore, $\chi |_{[\fh_{\ev},
\fh_{\ev}]\otimes\xi_1\xi_2}=0$.

Set
$$
\fm=\fh\otimes (\Lambda^1(\xi)\oplus\Lambda^2(\xi)).
$$

This is an ideal of
$\fg$ and $\fst_{\fn}(\chi)=\fm$. Therefore, the irreducible representation
$\rho$ of $\fm$ given by $\chi$ is such that $\rho|_{\fn}$ is a multiple of
$\chi$. Let
$$
\fn_1=\fh_{\ev}\otimes\Lambda^1(\xi)\oplus([\fh_{\ev},
\fh_{\ev}]\oplus \fh_{\od})\otimes\Lambda^2(\xi).
$$
This is an ideal in $\fm$ and
$$
[\fn_1, \fm]\subset([\fh_{\ev}, \fh_{\ev}]\oplus
\fh_{\od})\otimes\Lambda^2 (\xi)\subset \ker\; \chi.
$$

Thus, $\fn_1\subset \ker\;\rho$ since the subspace
$V^{\rho(\fn_1)}=\{v\in V:\;\rho(\fn_1) v=0\}$ is
$\rho(\fm_1)$-invariant and non-zero. But $\fm/\fn_1$ is a solvable 
Lie algebra and by the Lie theorem any its irreducible finite 
dimensional representation is of the same form as $\chi$. Let it be
$\chi$, for definiteness sake. 

This means that $\chi|_{[\fh_{\od},
\fh_{\od}]\otimes\Lambda^2(\xi)}=0$, i.e., $\chi|_{
\fh\otimes\Lambda^2(\xi)}=0$ and since $\fh\otimes\xi_1 \xi_2$ is an
ideal in $\fk$, then the representation of $\fk$ given by $\chi$
cannot be faithful. Lemma A1.2 is proved. \qed 

\ssec{A1.5. Central extensions of semisimple Lie superalgebras} 
Let $\fg$  be a Lie superalgebra, $\fr$ its radical of dimension
1. Then  $\fr$ is the center of $\fg$ and $\tilde\fg=\fg/\fr$ is semisimple.

First, consider the case when 
$\tilde\fg= \mathop{\oplus}\limits_{i\leq k}\fs_i$, where the $\fs_i$ are almost
simple and $k>1$. Let $\pi:\fg\rightarrow\tilde\fg$ be the natural projection. The
Lie superalgebra $\pi^{-1}(\fs_1)=\fk$ is an ideal in $\fg$ and $\dim \fk>1$. 

\ssbegin{A1.5.1}{Lemma} If $k>1$, then  $\rho$ can not
be irreducible of type A with respect to $\fk$.
\end{Lemma}

\begin{proof} Set $\fg_+=\mathop{\oplus}\limits_{i\geq 1}(\fs_i)_{\ev}$ and
$\fg_+$ is a Lie algebra. Then $[\fs_1, \fg_+ ]=0$. Since $\rho(\fr)$ acts by
scarlar operators and $\rho$ is a finite dimensional representation, it follows
that
$[\pi^{-1}(\fg_+), \fk]=0$, i.e., $\rho(\pi^{-1}(\fg_+))$ consists of
intertwining operators of $\rho|\fn$, i.e., $\rho(\pi^{-1} (\fg_+))=\Cee\cdot
1$ and, since $\rho$ is faithful, then $\pi^{-1} (\fg_+)=\fr$ implying
$\fg_+=0$. \end{proof}

\ssec{A1.5.2} Now, let $\tilde\fg$ contain an ideal $\fk$ of the form $(A1.1)$. The
central extension is defined by a cocycle $c: \tilde\fg\times
\tilde\fg\longrightarrow\Cee$. The cocycle condition is
$$
c(f, [g, h])=c([f, g], h)+(-1)^{p(f)p(g)}c(g, [f, h])\quad \text{for
any} f, g, h\in\tilde\fg.
$$
As earlier, we assume that $\fg$ has a faithful finitedimentional
representation; so the restriction of $c$ to $\tilde\fg_{\bar
0}\times \tilde\fg_{\bar 0}$ is trivial. Besides, $c|_{\tilde\fg_{\bar
0}\times \tilde\fg_{\bar 1}}=0$ by parity considerations. Therefore,
$$
c|_{\tilde\fg_{\bar
0}\times \tilde\fg}=0.\eqno{(A1.2)}
$$ 

\begin{th*}{Lemma} Set $\fL^m=\fh\otimes\Lambda^m(\xi_1, \dots ,
\xi_n)$. We have $c|_{\fL^{n}\times\fL^{n}}=0$.
\end{th*}

\begin{proof} If $n=2k+1$ the condition (A1.2) means that 
$$
c|_{\fh_{\bar 1}\otimes\Lambda^n(\xi)\times\fh_{\bar 1}\otimes\Lambda^n(\xi)}=0.
$$
Let $g_{\bar 1}, h_{\bar 1}\in\fh_{\bar 1}$ and $h_{\bar 0}\in\fh_{\bar
0}$. Then 
$$
\begin{matrix} 
c(f_{\bar0}\otimes\xi_1\cdot\dots\cdot\xi_n, [g_{\bar 1},
h_{\bar 1}]\otimes\xi_1\cdot\dots\cdot\xi_n)=
-c(f_{\bar 0}\otimes\xi_1\cdot\dots\cdot\xi_n, 
[g_{\bar 1}\otimes\xi_1\cdot\dots\cdot\xi_n, h_{\bar
1}])=\\
-c([f_{\bar 0}\otimes\xi_1\cdot\dots\cdot\xi_n, 
g_{\bar 1}\otimes\xi_1\cdot\dots\cdot\xi_n], h_{\bar
1})-c(g_{\bar 1}\otimes\xi_1\cdot\dots\cdot\xi_n, 
[f_{\bar 0}\otimes\xi_1\cdot\dots\cdot\xi_n, h_{\bar
1}])=0.\end{matrix} 
$$
Since $\fh_{\bar 0}=[\fh_{\bar 1}, \fh_{\bar 1}]$ (by Lemma A1.3), it
follows that $c|_{\fL^{n}\times\fL^{n}}=0$.

If $n=2k$ the condition (A1.2) means that 
$$
c|_{\fh_{\bar 0}\otimes\Lambda^n(\xi)\times\fh_{\bar 0}\otimes\Lambda^n(\xi)}=0.
$$
Let $f_{\bar 1}, h_{\bar 1}\in\fh_{\bar 1}$ and $g_{\bar 0}\in\fh_{\bar
0}$. Then 
$$
\begin{matrix} 
c(f_{\bar 1}\otimes\xi_1\cdot\dots\cdot\xi_n, [g_{\bar 0},
h_{\bar 1}]\otimes\xi_1\cdot\dots\cdot\xi_n)=
-c(f_{\bar 1}\otimes\xi_1\cdot\dots\cdot\xi_n, 
[g_{\bar 0}\otimes\xi_1\cdot\dots\cdot\xi_n, h_{\bar
1}])=\\
-c([f_{\bar 1}\otimes\xi_1\cdot\dots\cdot\xi_n, 
g_{\bar 0}\otimes\xi_1\cdot\dots\cdot\xi_n], h_{\bar
1})-c(g_{\bar 0}\otimes\xi_1\cdot\dots\cdot\xi_n, 
[f_{\bar 1}\otimes\xi_1\cdot\dots\cdot\xi_n, h_{\bar
1}])=0.\end{matrix} 
$$
Since $\fh_{\bar 1}=[\fh_{\bar 0}, \fh_{\bar 1}]$ (by Lemma A1.3), it
follows that
$c|_{\fL^{n}\times\fL^{n}}=0$. \end{proof}

\ssbegin{A1.5.3}{Lemma} $c|_{\fL^{k}\times\fL^{m}}=0$ for any $m=0$, 1,
\dots , $n$ and $k=n-m$, $n-m+1$, \dots, $n$.
\end{Lemma}

\begin{proof} We will perform the inverse double induction on $k$ and
$m$. For $k=m=n$ Lemma A1.5.2 will do.

Let the statement be true for all $k> k_0$ and $k_0+m\geq n$. Let
us show that it is true for $k=k_0$ as well. Observe that due to
sec. A2.6, $\tilde\fg$ contains $n$ elements $\eta _i$ such that $\ad\eta
_i|_{\fh}=\partial_{\xi_{i}}+D_i$, where $D_i\in\fvect(0|n)$ and $\deg
D_i>0$ in the natuaral grading of $\fvect$. Let $\varphi\in
\Lambda^{k_{0}+1}(\xi_1, \dots, \xi_n), \psi\in
\Lambda^{m}(\xi_1, \dots, \xi_n);\quad g, h\in\fh$ and $p(g\otimes
\psi)=\bar 0$; $p(h\otimes \psi)=\bar 1$.

Then we have 
$$
\begin{matrix} 
c([\eta_i, g\otimes\varphi],
h\otimes\psi)=(-1)^{p(g)}c(g\otimes\eta_i(\varphi), h\otimes\psi)\\
=
(-1)^{p(g)}\left(c(g\otimes\frac{\partial\varphi}{\partial\xi_{i}},
h\otimes\psi)+c(g\otimes D_{i}\varphi,
h\otimes\psi)\right).\end{matrix} 
$$
As to $D_i\varphi\in\mathop{\oplus}\limits_{s\geq k_{0}+1}\Lambda^s(\xi)$, the last
summand is equal to zero by the inductive hypothesis. 

On the other hand, 
$$
c([\eta_i, g\otimes\varphi],
h\otimes\psi)=c(\eta_i, [g\otimes\varphi,
h\otimes\psi])+c([\eta_i,  h\otimes\psi], g\otimes\varphi). 
$$
As $\deg\varphi+\deg\psi=k_0+1+m>n$, the bracket in the first summand
above is equal to 0. The second summand is equal to 0 by (A1.2). So
$c(g\otimes\frac{\partial\varphi}{\partial\xi_{i}},
h\otimes\psi)=0$, for arbitrary
$g\otimes\frac{\partial\varphi}{\partial\xi_{i}}\in\fL_{\bar 1}^{k_{0}}$
and  $h\otimes \psi \in\fL_{\bar
1}^{m}$. \end{proof}

\ssbegin{A1.5.4}{Lemma} $\fk$ has no faithful irreducible finite
dimensional representations.
\end{Lemma}

\begin{proof} Word-for-word proof of Lemmas A1.2.1 with the
help of Lemmas A1.5.2 and A1.5.3. \end{proof}

\begin{th*}{Corollary} $\rho$ can not be of type A with respect to the ideal
$\fk$.\end{th*}

\section*{Appendix 2: Background}

\ssec{A2.1. Linear algebra in superspaces}
A {\it superspace} is a $\Zee/2$-graded space; for a
superspace $V=V_{\bar 0}\oplus V_{\bar 1}$ denote by $\Pi (V)$ another copy of
the same superspace: with the shifted parity, i.e., $\Pi(V_{\bar i})= V_{\bar
i+\bar 1}$. 

A superspace structure in $V$ induces the superspace structure in the space
$\End (V)$. A {\it basis of a superspace} is always a basis consisting of {\it
homogeneous} vectors; let $\Par=(p_1, \dots, p_{\dim V})$ be an ordered
collection of their parities, called the {\it format} of $V$. 

A {\it supermatrix} of format (size) $\Par$ is a $\dim V\times \dim
V$ matrix whose $i$th row and column are said to be of parity $p_i$. The matrix
unit $E_{ij}$ is supposed to be of parity $p_i+p_j$ and the bracket of
supermatrices (of the same format) is defined via Sign Rule: 

\noindent{\it if something
of parity $p$ moves past something of parity $q$ the sign $(-1)^{pq}$ accrues;
the formulas defined on homogeneous elements are extended to arbitrary ones via
linearity}. 

For example: $[X, Y]=XY-(-1)^{p(X)p(Y)}YX$; the sign $\wedge$ of the
exterior or wedge product is also understood in this text in the supersence,
etc. 

Usually, $\Par$ is of the form $(\ev, \dots, \ev, \od, \dots, \od)$. Such a
format is called {\it standard}. 

The {\it general linear} Lie superalgebra of all supermatrices of size $\Par$ is
denoted by $\fgl(\Par)$, usually $\fgl(\ev, \dots, \ev, \od, \dots, \od)$ is
abbreviated to $\fgl(\dim V_{\ev}|\dim V_{\od})$.

The supercommutative superalgebra with unit generated by odd indeterminates
$\theta_1$, \dots ,$\theta_n$ is called Grassmann superalgebra and is denoted
by $\Lambda (n)$ or $\Lambda (\theta)$. The Lie superalgebra of
superderivations of $\Cee[x, \theta]$ is denoted by $\fvect(m|n)$. Clearly,
every element from $\fvect(m|n)$ is of the form
$$
D=\sum_{i=1}^mf_i(x, \theta)\pder{x_i}+\sum_{j=1}^ng_i(x,
\theta)\pder{\theta_j}.
$$

Define the {\it divergence} of a vector field setting
$$
\Div(\sum_{i=1}^mf_i(x, \theta)\pder{x_i}+\sum_{j=1}^ng_i(x,
\theta)\pder{\theta_j})=\sum_{i=1}^m\pderf{f_i}{x_i}+\sum_{j=1}^n(-1)^{p(g_{i})}
\pderf{g_i}{\theta_j}.
$$
The Lie superalgebra of divergence-free vector fields is denoted by
$\fsvect(m|n)$.

\ssec{A2.2. General linear superalgebras and irreducibilities: two types} The
Lie superalgebra
$\fgl(m|n)$ is called the {\it general Lie superalgebra}. Its supermatrices (in
the standard format) can be expressed as the sum of the even and odd parts: 
$$
\begin{pmatrix}A&B\\ C&D\end{pmatrix}=\begin{pmatrix}A&0\\
0&D\end{pmatrix}+\begin{pmatrix}0&B\\ C&0\end{pmatrix}.
$$

The {\it supertrace} is the map $\fgl (\Par)\longrightarrow \Cee$,
$(A_{ij})\mapsto \sum (-1)^{p_{i}}A_{ii}$. The space of supertraceless matrices
constitutes the {\it special linear} Lie subsuperalgebra $\fsl(\Par)$.

There are, however, two super versions of $\fgl(n)$, not one. The other version
is called the {\it qeer} Lie superalgebra  and is defined as the one that
preserves the complex structure given by an {\it odd} operator $J$, i.e., is
the centralizer $C(J)$ of $J$:
$$
\fq(n)=C(J)=\{X\in\fgl(n|n): [X, J]=0 \}, \text{ where } J^2=-\id.
$$
It is clear that by a change of basis we can reduce $J$ to the form
$J_{2n}=\begin{pmatrix}0&1_n\\ -1&0\end{pmatrix}$. In the standard format we have
$$
\fq(n)=\left \{\begin{pmatrix}A&B\\ B&A\end{pmatrix}\right\}.
$$
On $\fq(n)$, the {\it queer trace} is defined: $\qtr: \begin{pmatrix}A&B\\
B&A\end{pmatrix}\mapsto
\tr B$. Denote by $\fsq(n)$ the Lie superalgebra of {\it queertraceless}
matrices.

Observe that the identity representations of $\fq$ and $\fsq$ in $V$, though
irreducible in supersetting, are not irreducible in the nongraded sence: 
take homogeneous linearly independent vectors $v_1$, \dots , $v_n$ from $V$;
then $\Span (v_1+J(v_1), \dots , v_n+J(v_n))$ is an invariant subspace of $V$
which is not a subsuperspace.

A representation is called {\it irreducible of $G$-type} if it has no
invariant subspace; it is called {\it irreducible of $Q$-type} if it has no
invariant sub{\it super}space, but has an invariant subspace.

\ssec{A2.3. Lie superalgebras that preserve bilinear forms: two types} To the
linear map $F$ of superspaces there corresponds the dual map $F^*$ between the
dual superspaces; if $A$ is the supermatrix corresponding to $F$ in a format
$\Par$, then to $F^*$ the {\it supertransposed} matrix $A^{st}$ corresponds:
$$
(A^{st})_{ij}=(-1)^{(p_{i}+p_{j})(p_{i}+p(A))}A_{ji}.
$$

The supermatrices $X\in\fgl(\Par)$ such that 
$$
X^{st}B+(-1)^{p(X)p(B)}BX=0\quad \text{for a homogeneous matrix
$B\in\fgl(\Par)$} 
$$
constitute a Lie superalgebra $\faut (B)$ that preserves a homogeneous bilinear
form on $V$ with matrix $B$. 

Recall that the {\it supersymmetry} of the homogeneous
form $\omega $ means that its matrix $F$ satisfies the condition
$F^{u}=F$, where $F^{u}=
\begin{pmatrix} 
R^{t}_{n} & -T^{t} \\ -S^{t} & -U^{t}_{m}\end{pmatrix}$ 
for the matrix $F=\begin{pmatrix} R_{n} & S \\ T & U_{m}\end{pmatrix}$, where
the subscripts $n$ and $m$ indicate the size of the square matrices.

A nondegenerate supersymmetric even bilinear form can be reduced to a canonical
form whose matrix in the standard format is  
$$
B_{ev}(m|2n)= \begin{pmatrix}  
1_m&0\\
0&J_{2n}
\end{pmatrix},\quad \text{where }J_{2n}=\begin{pmatrix}0&1_n\\
-1_n&0\end{pmatrix},
$$
or
$$
B'_{ev}(m|2n)= \begin{pmatrix}  
\antidiag (1, \dots , 1)&0\\
0&J_{2n}
\end{pmatrix}. 
$$
The usual notation for $\faut (B_{ev}(\Par))$ in the standard format is
$\fosp(m|2n)=\fosp^{sy}(m|2n)$. (Observe that the passage from $V$ to $\Pi (V)$
sends the supersymmetric forms to superskew-symmetric ones, preserved by
$\fsp'\fo(2n|m)\cong\fosp^{sk}(m|2n)$.)

In the standard format the matrix realizations of these algebras
are: 
$$
\begin{matrix} 
\fosp (m|2n)=\left\{\left (\begin{matrix}  E&Y&-X^t\\
X&A&B\\
Y^t&C&-A^t\end{matrix} \right)\right\};\quad \fosp^{sk}(2n|m)=
\left\{\left(\begin{matrix} A&B&X\\
C&-A^t&Y^t\\
Y^t&-X^t&E\end{matrix} \right)\right\}, \\
\text{where}\; 
\left(\begin{matrix} A&B\\
C&-A^t\end{matrix} \right)\in \fsp(2n)\; \text{and}\; E\in\fo(m).\end{matrix}  
$$

A nongegenerate supersymmetric odd bilinear form $B_{odd}(n|n)$ can be reduced
to a canonical form whose matrix in the standard format is  $J_{2n}$. A
canonical form of the superskew odd nondegenerate form  in the standard format
is $\Pi_{2n}=\begin{pmatrix}   0&1_n\\1_n&0\end{pmatrix}$. The usual notation for $\faut
(B_{odd}(\Par))$ is $\fpe^{sy}(n)$. The passage from $V$ to $\Pi (V)$ sends the
supersymmetric forms to superskew-symmetric ones and establishes an isomorphism
$\fpe^{sy}(2n|m)\cong\fpe^{sk}(n)$. This Lie superalgebra is called
{\it periplectic}. The matrix realizations in the standard format of these
superalgebras is:
$$
\begin{matrix}\fpe^{sy}(n)=\left\{\begin{pmatrix}A&B\\
C&-A^t\end{pmatrix}, \; \text{where}\; B=B^t, C=-C^t;\right\};\\
\fpe ^{sk}\ (n)=\left\{\begin{pmatrix}  A&B\\
C&-A^t\end{pmatrix}, \; \text{where}\; B=-B^t, C=C^t\right\}.
\end{matrix}
$$

The {\it special periplectic} superalgebra is $\fspe(n)=\{X\in\fpe(n): \str
X=0\}$.

\ssec{A2.4. A. Sergeev's central extension} A.~Sergeev proved that there is just
one nontrivial central extension of
$\fspe(n)$. It exists only for $n=4$ and is denoted by $\fas$. Let
us represent an arbitrary element $A\in\fas$ as a pair $A=x+d\cdot z$, where
$x\in\fspe(4)$,
$d\in{\Cee}$ and $z$ is the central element. The bracket in $\fas$ in the
matrix form is 
$$
\left[\begin{pmatrix} a & b \cr 
c & -a^t \end{pmatrix}+d\cdot z, \begin{pmatrix}
a' & b' \cr 
c' & -a'{}^t \end{pmatrix} +d'\cdot z\right]=\left[\begin{pmatrix} a & b \cr 
c & -a^t \end{pmatrix}, \begin{pmatrix}
a' & b' \cr 
c' & -a'{}^t \end{pmatrix}\right]+\tr~cc'\cdot z.
$$

\ssec{A2.5. Projectivization}. If $\fs$ is a Lie algebra of scalar matrices, 
and $\fg\subset \fgl (n|n)$ is a Lie subsuperalgebra containing $\fs$, 
then the {\it projective} Lie superalgebra of type $\fg$  is $\fpg=
\fg/\fs$. Lie superalgebras $\fg_1\bigodot \fg_2$ described in Introduction
are also projective.

Projectivization sometimes leads to new Lie superalgebras, for example: $\fpgl
(n|n)$,  $\fpsl (n|n)$,  $\fpq (n)$,  $\fpsq (n)$; whereas
$\fpgl (p|q)\cong \fsl (p|q)$ if $p\neq q$.

\ssec{A2.6. Simplicity} The Lie superalgebras $\fsl(m|n)$ for $m> n\geq 1$,
$\fpsl(n|n)$ for $n>1$, $\fpsq(n)$ for $n>2$, $\fosp(m|2n)$ for $mn\neq 0$ and
$\fspe(n)$ for
$n>2$ are simple.

We say that $\fh$ is {\it  almost simple} if it can be included
(nonstrictly) between a simple Lie superalgebra $\fs$ and the Lie superalgebra
$\fder~\fs$ of derivations of the latter: $\fs\subset\fh\subset\fder~\fs$.

By definition $\fg$ is {\it semisimple} if its radical is zero. V. Kac
[K] described semisimple Lie superalgebras. Let $\fs_1$, ... , $\fs_k$ be
simple Lie superalgebras, let $n_1$, ... , $n_k$ be {\it pairs} of nonnegative
integers $n_j=(n_j^\ev, n_j^\od)$, $\cF(n_j)$ be the supercommutative superalgebra
of polynomials of $n_j^\ev$ even and $n_j^\od$ odd variables, and
$\fs=\oplus\fs_j\otimes\cF(n_j)$. Then
$\fder\fs=\oplus\left((der\fs_j)\otimes\cF(n_j)\supplus 1\otimes
\fvect(n_j)\right)$. Let
$\fg$ be a subalgebra of $\fder\fs$ containing
$\fs$. 

1) If the projection of $\fg$ on $1\otimes\fvect(n_j)_{-1}$ coincides with
$\fvect(n_j)_{-1}$ for each $j=1,
\dots , k$, then $\fg$ is semisimple.

2) All semisimple Lie superalgebras arise in the manner
indicated. For finite dimensional $\fg$ all the ingredients must be also of
finite dimension, e.g., $n_j^\ev=0$ for all $j$.

\ssbegin{A2.7}{Lemma} {\em 1)} Let $\fg$ be an irreducible (in $V$) finite
dimensional matrix Lie superalgebra. Then the space $\cB$ of $\fg$-invariant
bilinear forms on $V$ is at most 1-dimensional. 

{\em 2)} If $\dim \cB=1$ or $\varepsilon$, then $\fg$ is $G$-irreducible.
\end{Lemma}

\begin{proof} Since the kernel of any $\fg$-invariant form is a $\fg$-invariant
subspace, every $\fg$-invariant form is either nondegenerate or identically zero.

Let $\cB\neq 0$; let $B\in \cB$ be a nondegenerate form. We denote its Gram
matrix also by $B$. The invariance of $B$ means that
$$
X^{st}B+(-1)^{p(B)p(X)}BX=0\quad\text{ for any }\quad X\in\fg, \text{or}\quad 
X^{st}=-(-1)^{p(B)p(X)}BXB^{-1}.\eqno{(A2.1)}
$$
For another form $C\in\cB$ we have from (A2.1):
$$
(-1)^{p(B)p(X)}BXB^{-1}C+(-1)^{p(C)p(X)}CX=0
$$
implying that $X$ preserves the operator with matrix $B^{-1}C$. 

If $\fg$ is $G$-irreducible, this immideately implies that $B^{-1}C=c\cdot \id$,
i.e., $B=c\cdot C$ for a constant $c$. 

If $\fg$ is $Q$-irreducible, then $B^{-1}C\in\Span(\id, J)$; in particular,
$\dim\cB_\ev\leq 1$ and $\dim\cB_\od\leq 1$. Denote by $sign $ the operator
with matrix $\diag (1_n,-1_m)$ in the standard format; i.e., $\id$ on the even
subspace and $-\id$ on the odd one. Supertransposing formula (A2.1) and taking
into account that $(X^{st})^2=sign  \cdot X\cdot sign $, we see that the form
$B^{st}~sign $ is $\fg$-invariant. Having multiplied (A2.1) by $J^{st}$ from
the left and by $J$ from the right and taking into account that $[x, J]=0$ for
any $X\in\fg$, we see that the form $JBJ$ is also $\fg$-invariant. All the
three forms: $B$, $B^{st}~sign $ and $JBJ$ are of the same parity and,
therefore, are proportional.

If $p(B)=\ev$, then 
$$
B=\begin{pmatrix}B_0&0\\0&B_2\end{pmatrix},
\quad B^{st}~sign =\begin{pmatrix}B_0^t&0\\0&-B_2^t\end{pmatrix},
\quad JBJ=\begin{pmatrix}-B_2&0\\0&B_0\end{pmatrix}.
$$
If $B$ and $B^{st}~sign $ are proportional, then one of the matrices $B_0$,
$B_2$ is symmetric and the other one skew-symmetric. If $B$ and $JBJ$ are
proportional, then the matrices $B_0$ and
$B_2$ are proportional. This holds simultaneously only if $B_0=0$ and $B_2=0$.

If $p(B)=\od$, then 
$$
B=\begin{pmatrix}0&B_1\\B_3&0\end{pmatrix},
\quad B^{st}~sign =\begin{pmatrix}0&B_3^t\\B_1^t&0\end{pmatrix},
\quad JBJ=\begin{pmatrix}0&B_3\\-B_1&0\end{pmatrix}
$$
implying $B_1=B_3=0$.

Therefore, if $\fg$ is $Q$-irreducible, then $\cB=0$.

\end{proof}

\end{document}